\documentclass[cernpreprint, atlasdraft=false, texlive=2020, UKenglish, texmf]{atlasdoc}
 
\usepackage[subfigure]{atlaspackage}
\usepackage{atlaspackage}
\usepackage{adjustbox}
\usepackage{atlasbiblatex}
\usepackage{multirow}
\usepackage{bigstrut}
\usepackage{longtable}
\usepackage{paralist}
\usepackage{tabularx}

\usepackage{atlasphysics}
 
\graphicspath{{logos/}{figures/}}
 
\usepackage{ANA-HIGG-2018-09-PAPER-defs}

\AtlasTitle{Search for  heavy resonances decaying into a pair of $Z$ bosons in the $\ell^+\ell^-\ell'^+\ell'^-$ and  $\ell^+\ell^-\nu\bar\nu$ final states using 139 $\mathrm{fb}^{-1}$ of proton--proton collisions at $\sqrt{s} = 13\,$TeV with the ATLAS detector}
 
\AtlasAbstract{
A  search for heavy resonances decaying into a pair of $Z$ bosons leading to $\ell^+\ell^-\ell'^+\ell'^-$ and  $\ell^+\ell^-\nu\bar\nu$ final states, where $\ell$ stands for either an electron or a muon, is presented.
The search uses proton--proton collision data at a centre-of-mass energy of 13 TeV collected from 2015 to 2018
that corresponds to the integrated luminosity of 139 $\mathrm{fb}^{-1}$ recorded by the ATLAS detector during Run~2 of the Large Hadron Collider.
Different mass ranges spanning 200 GeV to 2000 GeV for the hypothetical resonances are considered, depending on the final state and model. In the absence of a significant observed excess, the results are interpreted as upper limits on the production cross section of a spin-0 or spin-2 resonance.  The upper limits for the spin-0 resonance are translated to exclusion contours in the context of Type-I and Type-II two-Higgs-doublet models, and the limits for the spin-2 resonance are used to constrain the Randall--Sundrum model with an extra dimension giving rise to spin-2 graviton excitations.}
 
\author{The ATLAS Collaboration}
 
\AtlasRefCode{HIGG-2018-09}
 
\AtlasJournal{EPJC}

\PreprintIdNumber{CERN-EP-2020-153}
 
\AtlasJournalRef{Eur. Phys. J. C 81 (2021) 332}
\AtlasDOI{10.1140/epjc/s10052-021-09013-y}
\hypersetup{pdftitle={ATLAS document},pdfauthor={The ATLAS Collaboration}}
 
\begin{document}
 
\maketitle
 
\tableofcontents

\section{Introduction}
 
The discovery of a scalar particle by the ATLAS and CMS collaborations~\cite{HIGG-2012-27,CMS-HIG-12-028} in 2012, with measured properties~\cite{HIGG-2014-06,CMS-HIG-14-009,HIGG-2013-01,Aad:2019mbh,Sirunyan:2018koj} consistent with those of the Standard Model (SM)~\cite{Glashow:1961tr, Weinberg:1967tq, Salam:1968rm} Higgs boson, was a major milestone in the understanding of electroweak symmetry breaking~\cite{Englert:1964et,Higgs:1964pj,Guralnik:1964eu}. One important question is whether the discovered particle is part of an extended scalar sector as postulated by various extensions to the Standard Model such as the two-Higgs-doublet model (2HDM)~\cite{Branco:2011iw}. These extensions predict additional Higgs~bosons, motivating searches in an extended mass range.
 
This paper reports on two searches for heavy resonances decaying into two SM $Z$ bosons,
encompassing the final states produced from the subsequent $\zztollll$ and $\zztollvv$ decays,
where $\ell$ stands for either an electron or a muon and $\nu$ stands for all three neutrino flavours.
The data employed were recorded by the ATLAS detector between 2015 and 2018 in  proton--proton collisions at $\sqrt{s}=13~\tev{}$ and correspond to an integrated luminosity of \lumi. The additional Higgs boson (spin-0 resonance), denoted by $H$ throughout this paper, is assumed to be produced mainly via gluon--gluon fusion (ggF) and vector-boson fusion (VBF) processes with  the ratio of the two production mechanisms unknown in the absence of a specific model. The results are interpreted separately for the ggF and VBF production modes, with events being classified into ggF- and VBF-enriched categories in both final states, as discussed in Sections~\ref{sec:h4lanal} and~\ref{sec:hllvvanal}. The searches cover a wide mass range from $200~\gev{}$ up to $2000~\gev{}$ and look for an excess in the distribution of the the four-lepton invariant mass, \mfl, for the \llll final state, and the transverse mass, \mt, for the \llvv final state, as the escaping neutrinos do not allow the full reconstruction of the final state.
This mass range is chosen based on the sensitivity of the analysis as determined by the selection criteria and the size of the data sample.
 
The transverse mass is defined as:
 
\begin{equation*}
\mt \equiv \sqrt{ \left[{\sqrt{\mz^{2} + \left(\pt^{\ell\ell}\right)^2} + \sqrt{\mz^{2} + \left( \met \right)^2}}\: \right]^2 - \left\lvert \vecptll  + \vecmet \right\rvert^2 },
\end{equation*}
 
where \mz is the mass of the \Zboson boson~\cite{Tanabashi:2018oca},
$\vecptll$ and $\vecmet$ are the transverse momentum of the lepton pair and the missing transverse momentum
with magnitudes of  $\pt^{\ell\ell}$ and $\met$, respectively.
In the absence of such an excess, limits on the production rate of different signal hypotheses are obtained from a simultaneous likelihood fit in the two final states.  The hypothesis of a heavy Higgs boson in the narrow-width approximation (NWA) is studied.
The upper limits on the production rate of a heavy Higgs boson are also translated into exclusion contours in the context of the two-Higgs-doublet model.
As several theoretical models favour non-negligible natural widths, large-width assumption (LWA) models~\cite{Branco:2011iw}, assuming widths of 1\%, 5\%, 10\% and 15\% of the resonance mass,
are examined only for ggF production, which dominates in many scenarios over the next-largest contribution (VBF) in the search range.
Results are also interpreted assuming the bulk Randall--Sundrum (RS) model~\cite{rsModel1999, Davoudiasl200043}
with a warped extra dimension giving rise to a spin-2 Kaluza--Klein (KK) excitation of the graviton \grav.
 
The main improvements relative to the previous search~\cite{HIGG-2016-19} are the following:
i) full LHC \RunTwo integrated luminosity is used;
ii) both analyses profit from improved lepton reconstruction and isolation selection to mitigate the impact of additional \textit{pp} interactions in the same or neighbouring bunch crossing (pile-up);
iii) the reconstruction of jets uses a particle-flow algorithm which combines measurements from the tracker and the calorimeter;
iv) the normalisation of the SM \ZZ background is derived from data rather than being estimated from SM predictions;
v) event classification targeting different production processes
is optimised using machine learning (ML) algorithms in the case of $\zztollll$  final state; vi) the \mt~distribution is used to search for signals in the VBF-enriched category in the case of the $\zztollvv$ final state, in addition to the use of \mt~in the ggF-enriched category;
and vii) the search range is extended to $2000~\gev{}$ in signal mass.
The improved analyses reduce the expected upper limit on the production cross section of an additional heavy resonance by up to 40\% in comparison with the previous published result scaled to the full \runb\ luminosity.
Results of a similar search from a subset of data collected at the LHC with $\sqrt{s}=13~\tev{}$ have been reported by the CMS Collaboration in Ref.~\cite{Sirunyan:2018qlb}.

The paper is organised as follows. A brief description of the ATLAS detector is given in Section~\ref{sec:detector}. In Section~\ref{sec:samples} the data and simulated samples are described. The object reconstruction is described in Section~\ref{sec:objectreco}. The analysis strategies for the $\llll$ and  $\llvv$  final states are described in Sections~\ref{sec:h4lanal} and~\ref{sec:hllvvanal}, respectively. Section~\ref{sec:syst} describes the systematic uncertainties, Section~\ref{sec:results} the final results, and Section~\ref{sec:inter} the interpretation of these results in the various models.
 
\section{ATLAS detector}
\label{sec:detector}
 
The ATLAS experiment is described in detail in Ref.~\cite{PERF-2007-01}. ATLAS is a multipurpose detector with a forward--backward symmetric cylindrical geometry and a solid-angle\footnote{The ATLAS experiment uses a right-handed coordinate system with its origin at the nominal interaction point (IP) in the centre of the detector and the $z$-axis along the beam pipe. The $x$-axis points from the IP to the centre of the LHC ring, and the $y$-axis points upward. Cylindrical coordinates $(r,\phi)$ are used in the transverse plane, $\phi$ being the azimuthal angle around the $z$-axis. The pseudorapidity is defined in terms of the polar angle $\theta$ as $\eta=-\ln\tan(\theta/2)$.} coverage of nearly 4$\pi$.
The inner tracking detector (ID), covering the region $|\eta| <$ 2.5, consists of a silicon pixel detector, a silicon microstrip detector, and a transition-radiation tracker. The innermost layer of the pixel detector, the insertable B-layer~\cite{Capeans:1291633}, was installed between \RunOne and \RunTwo of the LHC. The inner detector is surrounded by a thin superconducting solenoid providing a $2~\text{T}$ magnetic field, and by a finely segmented lead/liquid-argon (LAr) electromagnetic calorimeter covering the region $|\eta| <$ 3.2. A steel/scintillator-tile hadron calorimeter provides coverage in the central region $|\eta| <$ 1.7. The endcap and forward regions, covering the pseudorapidity range 1.5 $< |\eta| <$ 4.9, are instrumented with LAr electromagnetic and hadron calorimeters, with steel, copper, or tungsten as the absorber material.
A muon spectrometer (MS) system incorporating large superconducting toroidal air-core magnets surrounds the calorimeters. Three layers of precision wire chambers provide muon tracking in the range $|\eta| <$ 2.7, while dedicated fast chambers are used for triggering in the region $|\eta| <$ 2.4. The trigger system, composed of two stages, was upgraded~\cite{TRIG-PERF} before \RunTwo. The first stage, implemented with custom hardware, uses information from the calorimeters and muon chambers to select events from the 40~MHz bunch crossings at a maximum rate of 100~kHz. The second stage, called the high-level trigger (HLT), reduces the data acquisition rate to about 1~kHz on average. The HLT is  software-based and runs reconstruction algorithms similar to those used in the offline reconstruction.

 
\section{Data and simulation}
\label{sec:samples}
The proton--proton ($pp$) collision data used in these searches were collected
by the ATLAS detector at a centre-of-mass energy of $13~\tev{}$  with a 25~ns
bunch-spacing configuration from 2015 to 2018. The data are subjected to
quality requirements: if any relevant detector component was not operating
correctly during the period in which an event was recorded, the event is rejected.
The efficiency for recording good-quality data during \runb\ is 95.6\%~\cite{Aad:2019hmz}.
 
Simulated events are used to determine the signal acceptance and some of the background contributions.
The events produced by each Monte Carlo (MC) event generator were processed through the ATLAS detector
simulation~\cite{SOFT-2010-01} within the $\GEANT4$ framework~\cite{GEANT4}.
Additional inelastic $pp$ interactions
were overlaid on the simulated signal and background events.
The MC event generator used for pile-up is \PYTHIA{8.186}~\cite{Sjostrand:2007gs} with the A2 set of
tuned parameters~\cite{ATL-PHYS-PUB-2012-003} and the MSTW2008LO~\cite{Martin2009MSTW2008} parton distribution function (PDF) set.
The simulated events are weighted to reproduce the
observed distribution of the mean number of interactions per bunch crossing in data (\pu reweighting).
 
Heavy spin-0 resonance production was simulated using the \POWHEGBOX{v2}~\cite{Nason:2013ydw}
MC event generator. The gluon--gluon fusion and vector-boson fusion production modes
were simulated separately, with matrix elements calculated to
next-to-leading-order (NLO) accuracy in quantum chromodynamics (QCD). \POWHEGBOX was interfaced to \PYTHIA{8.212}~\cite{Sjostrand:2014zea}
for parton showering and hadronisation with the AZNLO set of tuned parameters~\cite{STDM-2012-23},
and for decaying the Higgs boson into the \htollll{} or \htollvv{} final states.
The event generator was interfaced to the \textsc{EvtGen}~v1.2.0 program~\cite{Lange:2001uf} for the simulation
of bottom and charm hadron decays.
The leading-order (LO) \textsc{CT10} PDF set~\cite{Lai:2010vv} was used for the hard-scattering process.
Events from ggF and VBF production were generated in the resonance mass range of $300~\gev{}$ to $2000~\gev{}$
in the NWA, using a step size of 100$~\gev{}$ up to 1000$~\gev{}$ and 200$~\gev{}$ above.
For the \llll final state, due to the sensitivity of the analysis at lower masses, events were also
generated for $\mres = 200~\gev{}$.
In addition, events from ggF heavy Higgs production with a width of 15\% of the
Higgs boson mass \mres were generated at NLO accuracy in QCD with \MGNLO~v2.3.2~\cite{md52014},
which was interfaced to \PYTHIA{8.210}
for parton showering and hadronisation with the A14 set of tuned parameters (A14 tune)~\cite{ATL-PHYS-PUB-2014-021},
and for decaying the Higgs boson into the two leptonic final states.
The properties of bottom and charm hadron decays were simulated by \textsc{EvtGen}~v1.2.0.
Events were generated in the resonance mass range of $400~\gev{}$ to $2000~\gev{}$ using a step size of
100 (200)$~\gev{}$ up to (above) 1000$~\gev{}$.
Similarly, events with a width of 5\% or 10\% of $\mres = 900~\gev{}$ were generated for validating
the analytic parametrisation of the \mfl distribution used in the \llll final state as described in
Section~\ref{sec:modeling4l}.
For the \llvv final state, a reweighting procedure as described in Section~\ref{sec:modelingllvv} is used on fully
simulated events to obtain the reconstructed \mt distribution at any value of mass and width tested.

Spin-2 Kaluza--Klein gravitons from the bulk Randall--Sundrum model~\cite{Davoudiasl200043,PhysRevD.76.036006} were generated
with \MGNLO at LO accuracy in QCD with
the NNPDF2.3 LO PDF set with $\alpha_\mathrm{s} = 0.130$~\cite{Ball:2014uwa},
which is then interfaced to \PYTHIA{8.210} for parton showering and hadronisation
with the A14 tune and for decaying the heavy \ZZ resonance into the two leptonic final states.
The properties of bottom and charm hadron decays were simulated by \textsc{EvtGen}~v1.2.0.
The dimensionless coupling $k/{\overline{M}_\mathrm{Pl}}$, where $\overline{M}_\mathrm{Pl} = M_\mathrm{Pl}/\sqrt{8\pi}$
is the reduced Planck scale and $k$ is the curvature scale of the extra dimension, is set to 1.
The width of the resonance is correlated with the coupling $k/{\overline{M}_\mathrm{Pl}}$
and in this configuration it is around $\sim 6\%$ of its mass.
Mass points between $600~\gev{}$ and $2~\tev{}$ with $200~\gev{}$ spacing were generated for both final states.
 
The \qqZZ background was simulated by the
\SHERPA{v2.2.2}~\cite{Bothmann:2019yzt} generator,
in which the NNPDF3.0 NNLO PDF set~\cite{Ball:2014uwa} was used for the hard-scattering process,
achieving  NLO accuracy in the matrix-element calculation
for 0- and 1-jet final states and LO accuracy for 2- and 3-jet final states with
the \textsc{Comix}~\cite{Gleisberg:2008fv} and \textsc{OpenLoops}~\cite{Buccioni:2019sur,Cascioli:2011va,Denner:2016kdg}
matrix-element generators.
The merging with the \SHERPA parton shower~\cite{Schumann:2007mg} was
performed using the MEPS@NLO prescription~\cite{Hoeche:2012yf}.
NLO electroweak (EW) corrections were applied
as a function of \mfl\ for the \llll final state~\cite{Biedermann:2016yvs,Biedermann:2016lvg},
and as a function of the transverse momentum of the $Z$ boson that decays into two neutrinos
for the \llvv final state~\cite{Kallweit:2014xda, Kallweit:2017khh, Schonherr:2017qcj, Buccioni:2019sur, Kallweit:2019zez}.
The EW production of a \itzz\ pair and two additional jets via vector-boson scattering
up to $O(\alpha_\mathrm{EW}^6)$ was generated using \SHERPA{v2.2.2} for both the \llll~and \llvv final states,
where the process $ZZZ\rightarrow 4\ell qq$ is also taken into account.
In addition, the \WZ diboson events from both QCD and EW production,
with the subsequent leptonic decays of both the $W$ and $Z$ bosons, were simulated by \SHERPA with a similar set-up.
The \WZ~events with $Z$ boson decaying leptonically and $W$ boson decaying hadronically were modelled with \SHERPA{v2.2.1}.
 
The \ggZZ\ process was modelled by \SHERPA{v2.2.2} at LO accuracy in QCD for both final states,
including the off-shell SM \SMH boson contribution and the interference between the \SMH and \ZZ processes.
The higher-order correction factor accounting for up to NLO accuracy in QCD
for the \ggZZ continuum production was
calculated for massless quark loops~\cite{Caola:2015psa,Campbell:2016ivq, Alioli:2016xab} in the
heavy-top-quark approximation~\cite{Melnikov:2015laa}, including the $gg\rightarrow
h^{*} \rightarrow ZZ$ process~\cite{Li:2015jva}.
Based on these studies, a constant factor of 1.7
is used, and a relative uncertainty of 60\% is assigned to the normalisation in both searches.
 
For the \llvv~final state,
the contribution from $WW$ production was removed in the \SHERPA simulation of the \qqZZ~and \ggZZ~processes
by requiring the charged leptons and the neutrinos to have different lepton flavours.
The $q\bar{q}\rightarrow WW$ and $gg\rightarrow WW$ processes were then modelled with \POWHEGBOX{v2} and \SHERPA{v2.2.2}, respectively.
The interference between $WW$ and $ZZ$ production is expected to be negligible~\cite{Kallweit:2017khh} and is therefore not considered.

Events containing a single $Z$ boson with associated jets were simulated using the \SHERPA{v2.2.1} event generator.
Matrix elements were calculated for up to two partons at NLO and four partons at LO using
the \textsc{Comix} and \textsc{OpenLoops} matrix-element
generators and merged with the \SHERPA parton shower using
the MEPS@NLO prescription.
The NNPDF3.0 NNLO PDF set was used in conjunction
with dedicated parton-shower tuning developed by the \SHERPA authors. The \Zjets
events are normalised using the NNLO cross sections~\cite{GAVIN20112388}.
 
The triboson backgrounds $ZZZ$, $WZZ$, and $WWZ$ with fully leptonic decays and at least four prompt charged leptons were modelled
using \SHERPA{v2.2.2} with LO accuracy of the QCD calculations and the CT10 PDF set.
The simulation of $t\bar{t}+V$ production ($V =$ $W$ or $Z$) with both top quarks decaying semileptonically
and the vector boson decaying inclusively was performed with \MGNLO
interfaced to \PYTHIA{8.210} for parton showering and hadronisation
with the A14 tune
and to \textsc{EvtGen}~v1.2.0 for the simulation of bottom and charm hadron decays.
The total cross section is normalised to the prediction of Ref.~\cite{Frixione:2015zaa},
which includes the two dominant terms at both LO and NLO in a mixed perturbative expansion
in the QCD and EW couplings.
The $t\bar{t}$ background, as well as single-top and  $Wt$ production,
were modelled using \POWHEGBOX{v2} interfaced to \PYTHIA{8.230}
with the A14 tune
and to \textsc{EvtGen}~v1.6.0 for the simulation of bottom and charm hadron decays.
 
In order to study the interference treatment for the LWA case,
samples containing the \ggZZ\ continuum background (\textit{B})
as well as its interference (\textit{I}) with a hypothetical heavy Higgs signal (\textit{S}) were used
and are referred to as \textit{SBI} samples hereafter.
In the \llll  final state the \textsc{MCFM} NLO event generator~\cite{Campbell2011},
interfaced to \PYTHIA{8.212}, was used to produce \textit{SBI} samples
where the width of the heavy scalar is set to 15\% of its mass,
for masses of 200, 300, 400, 500, 600, 800, 1000, 1200, and 1400$~\gev{}$.
Background-only samples were also generated with the \textsc{MCFM} event generator,
and are used to extract the signal-plus-interference term (\textit{SI})
by subtracting them from the aforementioned \textit{SBI} samples.
For the \llvv final state, the  \textit{SBI} samples were generated with the
\textsc{gg2VV} event generator~\cite{Kauer:2012hd, Kauer:2013qba}.
The samples include signal events with a scalar mass of 400, 700, 900, 1200, and 1500$~\gev{}$.
 
\section{Event reconstruction}
\label{sec:objectreco}
Electron reconstruction uses a dynamic, topological calorimeter-cell clustering-based approach which allows improved measurement of the electron energy, particularly in situations where an electron radiates a bremsstrahlung photon; details can be found in Ref.~\cite{EGAM-2018-01}. Electron candidates are clusters of energy deposits in the calorimeter associated with ID tracks, where the final track--cluster matching is performed after the tracks have been fitted with a Gaussian-sum filter (GSF)~\cite{ATLAS-CONF-2012-047}  to account for bremsstrahlung energy losses.
The electron's transverse momentum is computed from the cluster energy and the track direction at the interaction point.
Background rejection relies on the longitudinal and transverse shapes of the electromagnetic showers in the calorimeters, track--cluster matching, and properties of tracks in the ID. All of this information, except for that related to track hits, is combined into a likelihood discriminant. The selection  combines the likelihood with the number of track hits and defines several working points (WP). Selected electrons have $\pt > 4.5~\gev{}$~and $|\eta| < 2.47$. The $\llll$ analysis uses a `loose' WP, with an efficiency of at least 90\% for electrons with $\pt>30~\gev{}$~\cite{PERF-2017-01}. The `medium' WP (with an efficiency about 85\% for electrons with $\pt>30~\gev{}$) is adopted to select candidate electrons in the \llvv~analysis.

Muons are formed from tracks reconstructed in the ID and MS, and their identification is primarily based on the presence of the track or track segment in the MS~\cite{PERF-2015-10}. If a complete track is present in both the ID and the MS, a combined muon track is formed by a global fit using the hit information from both the ID and MS detectors (combined muon); otherwise the momentum is measured using the ID, and the MS track segment serves as identification (segment-tagged muon). The segment-tagged muon is limited to the centre of the barrel region ($|\eta| <0.1$) which has reduced MS geometrical coverage. Furthermore, in this central region an ID track with $\pt > 15~\gev{}$ is identified as a muon if its calorimetric energy deposition is consistent with a minimum-ionising particle (calorimeter-tagged muon). In the forward region ($2.5 < |\eta| < 2.7$) with limited or no ID coverage, the MS track formed out of three MS layers is either used alone (stand-alone muon) or combined with silicon-detector hits, if found in the forward ID (combined muon). The ID tracks associated with the muons are required to have at least a minimum number of associated hits in each of the ID subdetectors to ensure good track reconstruction. The minimum \pt~for muon candidates is $5~\gev{}$, while the maximum $|\eta|$ is 2.7. A `loose' muon identification WP, which uses all muon types, is adopted by the  $\llll$ analysis.
This criterion has an efficiency of at least 98\%~\cite{PERF-2015-10} for isolated muons with $\pt = 5~\gev{}$ and rises to 99.5\% at higher \pt.  For the $\llvv$ analysis a `medium' WP is used, which only includes combined muons and has an efficiency of 98\%.

The reconstruction of jets uses a particle-flow algorithm~\cite{Aaboud:2017aca} which combines measurements from both the tracker and the calorimeter. The energy deposited in the calorimeter by all charged particles is removed, and the jet reconstruction is performed on an ensemble of `particle-flow objects' consisting of the remaining calorimeter energy and tracks which are matched to the hard interaction. This improves the accuracy of the charged-hadron measurement, while retaining the calorimeter measurements of neutral-particle energies. Compared to only using topological clusters~\cite{Aad:2016upy}, jets reconstructed with the particle-flow algorithm with \pt of about $30~\gev{}$ have approximately 10\% better transverse momentum resolution. The two different algorithms have similar resolutions for \pt above $100~\gev{}$. Particle-flow jets are reconstructed using the anti-$k_t$ algorithm~\cite{Cacciari:2008gp} with a radius parameter $R$ = 0.4.
The jet four-momentum is corrected for the calorimeter’s non-compensating response, signal losses due to noise threshold effects, energy lost in non-instrumented regions, and contributions from pile-up~\cite{Aaboud:2017jcu}.  The jets used are required to satisfy $\pt > 30~\gev{}$ and $|\eta |<4.5$. Jets from pile-up with $|\eta |<2.5$ are suppressed using a jet-vertex-tagger multivariate discriminant~\cite{ATLAS-CONF-2014-018,Aad:2015ina}.

Jets containing $b$-hadrons, referred to as $b$-jets, are identified by the long lifetime, high mass, and decay multiplicity of $b$-hadrons, as well as the hard $b$-quark fragmentation function.
The $\llvv$ analysis identifies $b$-jets of $\pt > 20~\gev{}$ and $|\eta |<2.5$ using an algorithm that achieves an identification efficiency of about 85\% in simulated $t\bar{t}$ events, with a rejection factor for light-flavour jets of about 30~\cite{Aad:2019aic}.
 
Selected events are required to have at least one vertex having at least two associated tracks with $\pt > 500~\mev{}$, and the primary vertex is chosen to be the vertex reconstructed with the largest $\sum \pt^2$ of its associated tracks.
As lepton and jet candidates can be reconstructed from the same detector information, a procedure to resolve overlap ambiguities is applied.
In the \llll case, the overlap ambiguities are resolved as follows. If two electrons have overlapping energy deposits,
the electron with the higher \pt is retained. If a reconstructed electron and muon share the same ID track, the muon is rejected if it is calorimeter-tagged; otherwise the electron is rejected. Reconstructed jets geometrically overlapping in a cone of size $\Delta R = 0.2$ with electrons or muons are also removed.
The overlap removal in the \llvv~case is similar to that in the \llll~case, except for an additional criterion that removes any leptons close to the remaining jets with $0.2 < \Delta R < 0.4$.
This additional criterion is not imposed in the \llll~case due to the cleaner environment of this final state
and in order to maximise the signal efficiency.

The missing transverse momentum \vecmet, which accounts for the imbalance of visible momenta in the plane transverse to the beam axis, is computed as the negative vector sum of the transverse momenta of all identified electrons, muons and jets, as well as a `soft term', accounting for unclassified soft tracks and energy clusters in the calorimeters~\cite{Aaboud:2018tkc}.
This analysis uses a track-based soft term, which is built by combining the information provided by the ID and the calorimeter, in order to minimise the effect of \pu, which degrades the \met resolution.
 
\section{Analysis of \llll~final state}
\label{sec:h4lanal}
\subsection{Event selection and categorisation}
 
In Section~\ref{sec:commonsel4l} the  four-lepton event selection is described. After this selection,
events are further split into several categories, in order to probe different signal production modes, such as VBF production and ggF production.
To enhance the search sensitivity to the NWA signals, multivariate classifiers are optimised for the event categorisation as described in Section~\ref{sec:mvasel4l}.
In order to also obtain results that are more model-independent (since the training of the multivariate classifiers is usually based on a specific signal model), a cut-based event categorisation that enhances the sensitivity in the VBF production mode is also considered and is described in Section~\ref{sec:vbfsel4l}.
 
In the search for LWA signals, due to the complexity of modelling the categorisation of the interference between heavy Higgs boson and SM Higgs boson processes, only the \ggfenr categories of the cut-based analysis (CBA) are used.  The same strategy is adopted in the search for a Kaluza--Klein graviton excitation.

\subsubsection{Common event selection}
\label{sec:commonsel4l}
Four-lepton events are selected and initially classified according to the lepton flavours: $4\mu$, $2e2\mu$, $4e$, called `channels' hereafter.  They are selected using a combination of single-lepton, dilepton and trilepton triggers with different transverse momentum thresholds. The single-lepton triggers with the lowest \pt thresholds had tighter requirements than the high \pt threshold single-lepton triggers and the multilepton triggers.
Due to an increasing peak luminosity, these \pt thresholds increased during the data-taking periods~\cite{ATL-DAQ-PUB-2016-001, ATL-DAQ-PUB-2017-001}. For single-muon triggers, the \pt threshold increased from 20$~\gev{}$ to $26~\gev{}$, while for single-electron triggers, the \pt threshold increased from 24$~\gev{}$ to $26~\gev{}$. The overall trigger efficiency for signal events passing the final selection requirements is about 98\%.
 
In each channel, four-lepton candidates are formed by selecting a lepton-quadruplet made out of two same-flavour, opposite-sign lepton pairs, selected as described in Section~\ref{sec:objectreco}. Each electron (muon) must satisfy
$\pt > 7$ (5)$~\gev{}$ and be measured in the pseudorapidity range of
$\left|\eta \right|<2.47$ ($2.7$). The highest-\pt\ lepton in the quadruplet
must satisfy $\pt > 20~\gev{}$, and the second (third) lepton in \pt\ order must satisfy
$\pt > 15~\gev{}$ ($10~\gev{}$). In the case of muons, at most one calorimeter-tagged,
segment-tagged or stand-alone ($2.5 < |\eta| < 2.7$) muon is allowed per quadruplet.
 
If there is ambiguity in assigning leptons to a pair, only one quadruplet per channel is selected by keeping the quadruplet with the invariant mass of the lepton pairs closest (leading pair) and second closest (subleading pair) to  the $Z$ boson mass~\cite{Tanabashi:2018oca}, with invariant masses referred to as  $m_{12}$ and $m_{34}$ respectively.
In the selected quadruplet, $m_{12}$ must satisfy $50~\gev{} < m_{12} < 106~\gev{}$ and $m_{34}$ must satisfy $50~\gev{} < m_{34} < 115~\gev{}$.
 
Selected quadruplets are required to have their leptons separated from each other by $\Delta R>0.1$.
For $4\mu$ and $4e$ quadruplets, if an opposite-charge same-flavour lepton pair is found with $m_{\ell\ell}$ below $5~\gev{}$, the quadruplet is removed to suppress the contamination from $J/\psi$ mesons. If multiple quadruplets from different channels are selected at this point, only the quadruplet from the channel with the highest signal acceptance is retained, in the order: $4\mu$, $2e2\mu$, $4e$.
 
The \Zjets and \ttbar background contributions are reduced by imposing impact-parameter requirements as well as track- and calorimeter-based isolation requirements on the leptons. The transverse impact-parameter significance, defined as the impact parameter calculated relative to the measured beam-line position in the transverse plane divided by its uncertainty, $|d_0|/\sigma_{d_0}$, for all muons (electrons), is required to be lower than 3 (5).
The track-isolation discriminant is calculated  from the tracks with $\pt > 500~\mev{}$ that lie within a cone of $\Delta R=0.3$ around the muon or electron and that either originate from the primary vertex or have a longitudinal impact parameter $z_0$ satisfying $|z_0\sin(\theta)| < 3$~mm  if not associated with any vertex. Above a lepton \pt of $33~\gev{}$, the cone size falls linearly with \pt to a minimum cone size of 0.2 at $50~\gev{}$. Similarly, the calorimeter isolation is calculated from the positive-energy topological clusters that are not associated with a lepton track in a cone of $\Delta R=0.2$  around the muon or electron.  The sum of the track isolation and 40\% of the calorimeter isolation is required to be less than 16\% of the lepton \pt. The calorimeter isolation is corrected for electron shower leakage, pile-up, and underlying-event contributions. Both isolations are corrected for track and topological cluster contributions from the remaining three leptons. The pile-up dependence of this isolation selection is reduced compared with that of the previous search by optimising the criteria used for exclusion of tracks associated with a vertex other than the primary vertex and by the removal of topological clusters associated with tracks.

An additional requirement based on a vertex-reconstruction algorithm, which fits  the four-lepton candidates with the constraint that they originate from a common vertex, is applied in order to further reduce the  $Z+\mathrm{jets}$ and $t\bar{t}$ background contributions. 
A cut of $\chi^{2} / \mathrm{ndof} < 6$ for $4\mu$ and $< 9$ for the other channels is applied,
with an efficiency larger than  $99 \%$ for signal in all channels.
 
The QED process of radiative photon production in $Z$ boson decays is well modelled by simulation.
Some of the final-state-radiation (FSR) photons can be identified in the calorimeter and incorporated into
the \llll\ analysis.  The strategy to include FSR photons into the reconstruction of $Z$ bosons is
the same as in \runa~\cite{HIGG-2013-20}. It consists of a search for collinear (for muons)
and non-collinear FSR photons (for muons and electrons) with only one FSR photon allowed per event.
After the FSR correction, the four-momenta of both dilepton pairs are recomputed by
means of a $Z$-mass-constrained kinematic fit~\cite{HIGG-2013-12}. The fit uses a Breit--Wigner $Z$ boson lineshape and
a single Gaussian function  per lepton to model the momentum response function with the Gaussian width
set to the expected resolution for each lepton. The $Z$-mass constraint is applied to both $Z$
candidates.
 
Events that pass the common event selection (as described above) which are not yet split according to lepton flavours, form a category which is called `inclusive' hereafter.
 
\subsubsection{Event categorisation: multivariate analysis}
\label{sec:mvasel4l}
In order to improve the sensitivity in the search for an NWA Higgs boson signal produced either in the VBF or the ggF production mode,
two multivariate classifiers, namely a `VBF classifier' and a `ggF classifier', are used.
These classifiers are built with deep neural networks (DNN) and use a architecture similar to that in Ref.~\cite{Aad:2020mkp}, combining a multilayer perceptron (MLP) and one or two recurrent neural networks (rNN)~\cite{tensorflow2015-whitepaper}.
For both classifiers, the outputs of the MLP and rNN(s) are concatenated
and fed into an additional MLP that produces an event score.

The `VBF classifier' uses two rNNs and an MLP. The two rNNs have as inputs the \pt-ordered transverse momenta and the pseudorapidities of the two leading jets and the transverse momenta and the pseudorapidities of the four leptons in the event. The MLP uses as inputs the invariant mass of the four-lepton system, the invariant mass and the transverse momentum of the two-leading-jets system, the difference in pseudorapidity between the \llll system and the leading jet, and the minimum angular separation between the $\ell^+\ell^-$ or $\ell'^+\ell'^-$ pair and a jet.
 
The `ggF classifier' uses one rNN and an MLP. The rNN has as inputs the \pt-ordered transverse momenta and the pseudorapidities of the four leptons in the event.
The MLP uses as inputs the following variables:
\begin{inparaenum}[1)]
\item the four-lepton invariant mass;
\item the transverse momentum and the pseudorapidity of the four-lepton system;
\item the production angle of the leading \Z defined in the four-lepton rest frame, $\cos\theta^*$;
\item the angle between the negative final-state lepton and the direction of flight of leading (subleading) \Z in the \Z rest frame, $\cos\theta_1$ ($\cos\theta_2$);
\item the angle between the decay planes of the four final-state leptons expressed in the four-lepton rest frame, $\Phi$; and
\item the transverse momentum and the pseudorapidity of the leading jet.
\end{inparaenum}
 
\begin{figure}[!htbp]
\subfigure[]{\includegraphics[width=0.45\textwidth]{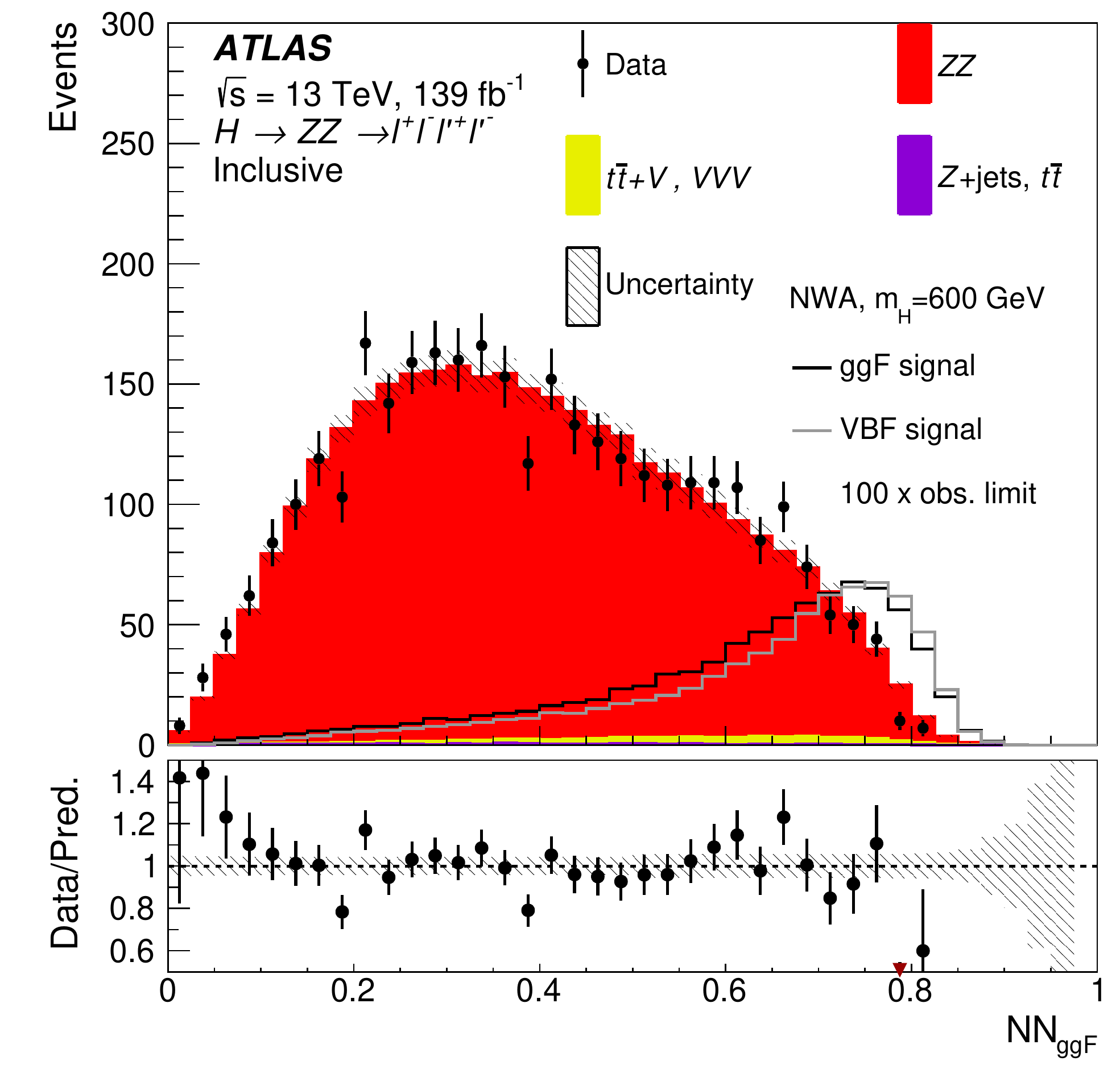}}
\subfigure[]{\includegraphics[width=0.45\textwidth]{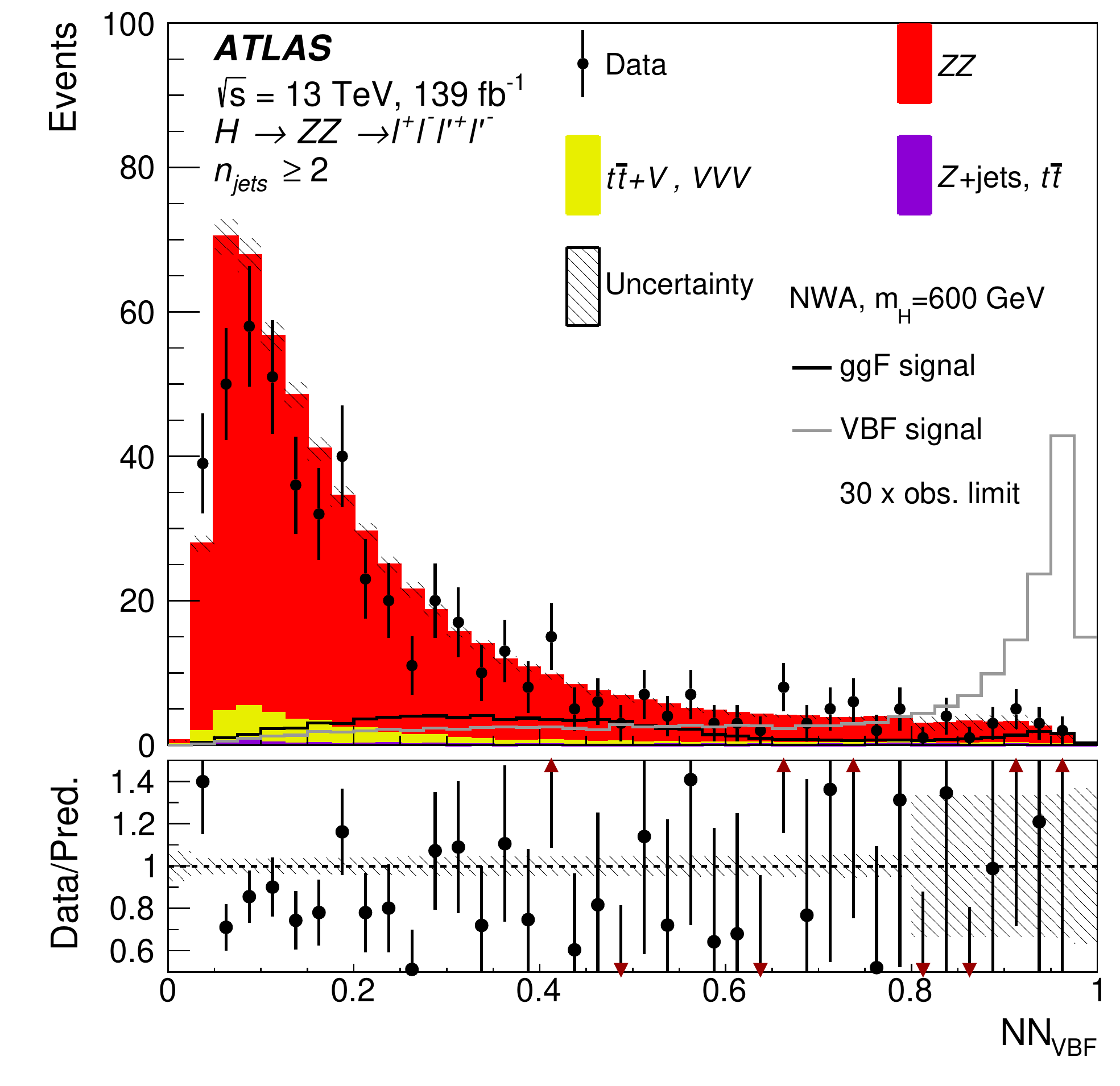}}
\caption{The output of (a) the `ggF classifier' (NN$\ensuremath_{\text{ggF}}$)
and (b) the `VBF classifier' (NN$\ensuremath_{\text{VBF}}$) for the events passing the common event selections
for the data, the SM background and NWA signal events with a mass of $600~\gev{}$.
For (b) the `VBF classifier' output, an additional requirement of at least two jets in the event, is applied.
The signal cross section is set to 100 times the observed limit for the `ggF classifier'
and 30 times the observed limit for the `VBF classifier'.
The \ZZ background is scaled by the normalisation factors shown in Table 2.
The lower panels show the ratio of data to prediction.
Only statistical and experimental systematic uncertainties are included.
}
\label{fig:NN_SR}
\end{figure}

The two classifiers are trained separately using the above-listed discriminating variables on all simulated NWA signal events from their corresponding production mode, and the SM \ZZ background events. The `VBF classifier' is trained on events with at least two jets while the `ggF classifier' is trained on  events with fewer than two jets. In order to represent the relative importance of the signal and background events,
weights that scale the events to the same luminosity according to their production cross sections are used in the training.  Furthermore, in order to achieve good discriminating power of the classifiers over a large range of signal mass hypotheses, the signal events are reweighted such that their overall four-lepton invariant mass spectrum matches that of the SM background events.
As a result of this reweighting method the classifiers do not produce a bias towards a specific mass point.
Extensive checks are performed to ensure such treatment does not create a local excess of background events that would fake a signal.
Figure~\ref{fig:NN_SR} shows the `ggF classifier' and  `VBF classifier' output for the data,
the SM background and an example signal with $\mH = 600~\gev{}$.
 
After the common event selection, as described in Section~\ref{sec:commonsel4l}, events with at least two jets ($n_{\text{jets}}\geq 2$) and a `VBF classifier' score value greater than 0.8 form the \vbfMVAenr category. Events failing to enter the \vbfMVAenr category are classified into the \ggfhigh category if the `ggF classifier' score value is greater than 0.5; these events are further split into three distinct categories according to the lepton flavour of the \llll system. Finally, events failing both classifiers form the ggF-low category. Overall, five mutually  exclusive categories are formed: \vbfMVAenr, \ggfhighfmu,  \ggfhighemu, \ggfhighfe,  \ggflow. This categorisation is used in the search for a heavy scalar with the NWA and in the search in the context of a CP-conserving 2HDM.
 
The signal acceptance, defined as the ratio of the number of reconstructed events after all selection requirements to the total number of simulated events, is found to be between 30\% (15\%) and 46\% (22\%) in the ggF (VBF)-enriched category for the ggF (VBF) production mode depending on the signal mass hypothesis.

\subsubsection{Event categorisation: cut-based analysis}
\label{sec:vbfsel4l}
As in the previous publication~\cite{HIGG-2016-19}, a cut-based analysis is also performed to probe the sensitivity in the VBF production mode. If an event has two or more jets with \pt greater than $30~\gev{}$, with the two leading  jets being well separated in $\eta$, $\detajj > 3.3$, and having an invariant mass $\mjj > 400~\gev{}$, this event is classified into the VBF-enriched category; otherwise the event is classified into one of the ggF-enriched categories further split according to the lepton flavour of the \llll system.
Four distinct categories are formed, namely \vbfCBAenr, \ggfCBAhighfmu, \ggfCBAhighemu, and \ggfCBAhighfe.
The \ggfenr categories are used in the search for a heavy large-width scalar and the search for a Kaluza--Klein graviton excitation. In addition, as for the multivariate-based analysis, such categorisation is used in the search for a heavy scalar with the NWA and the corresponding results are described in the Appendix.
 
\subsection{Background estimation}
\label{sec:bkg4l}
 
The main background source in the \htollll\ final state is non-resonant SM \itzz\ production,
accounting for 97\% of the total background events in the inclusive category.
It arises from quark–antiquark annihilation \qqZZ (86\%),
gluon-initiated production \ggZZ (10\%), and a small contribution from EW vector-boson scattering (1\%).
The last of these is more important in the VBF-enriched category using
the DNN-based categorisation, where it accounts for 20\% of the total background events.
While in the previous publication~\cite{HIGG-2016-19} the SM \itzz\ background was exclusively estimated from simulation for both the shape and the normalisation,
in this analysis its normalisation is derived from the data in the likelihood fit used in the statistical treatment of the data as explained in Section~\ref{sec:results}.
The shapes of the \qqZZ and \ggZZ invariant mass distributions are parameterised with analytic functions as described in
Section~\ref{sec:modeling4l}. Additional background comes from the \Zjets and \ttbar processes. These contribute to the total background yields at the percent level and decrease more rapidly than the non-resonant \itzz\ contribution as a function of \mfl. These backgrounds are estimated using data where possible, following slightly different approaches for final states with a dimuon (\llmumu) or a dielectron (\llee) subleading pair~\cite{HIGG-2016-25,HIGG-2016-22}.
 
The \llmumu\ non-\ZZ background comprises mostly \ttbar and \Zjets events, where in the latter case the muons arise mostly from heavy-flavour semileptonic decays and to a lesser extent from $\pi$/$K$ in-flight decays.
The normalisations of the \Zjets\ and \ttbar\ backgrounds are determined
by fitting the invariant mass of the leading lepton pair in dedicated data control regions.
The control regions are formed by relaxing the $\chi^2$ requirement on the four-lepton vertex fit,
and by inverting and relaxing isolation and/or impact-parameter requirements on the subleading muon pair.
An additional control region ($e\mu\mu\mu$) is used to improve the \ttbar\ background estimate.
The contribution of transfer factors, defined as the number of events in the signal region divided by the number of events in the control region, are obtained separately for \ttbar\ and \Zjets\ using simulated events to extrapolate the yields from the control regions to the signal regions.
 
The main non-prompt background for the \llee\ process arises from three sources: light-flavour jets misidentified as electrons; photon conversions; and semileptonic decays of heavy-flavour hadrons.  The \llee\ control-region selection requires the electrons in the subleading lepton pair to have the same charge, and relaxes the identification and isolation requirements on the electron candidate, denoted $X$, with the lower transverse momentum.
The heavy-flavour background is found to be negligible,
whereas the light-flavour and photon-conversion background is obtained with the sPlot~\cite{Pivk2005356} method, based on a fit to the number of hits in the innermost ID layer in the data control region.  Transfer factors for the light-flavour jets and converted photons, obtained from simulated samples, are corrected using a $Z+X$ control region and then used to extrapolate the extracted yields to the signal region.  Both the yield extraction and the extrapolation are performed in bins of the transverse momentum of the electron candidate and the jet multiplicity.
 
The $WZ$ production process is included in the data-driven estimates for the \llee\ final states,
while it is added from simulation for the \llmumu\ final states
even though its contribution to the total background is at the per-mille level.
The contributions from \ttv\ (where $V$ stands for either a $W$ or a $Z$ boson) and triboson
processes are minor and taken from simulated samples.

\subsection{Signal and background modelling}
\label{sec:modeling4l}
The reconstructed four-lepton invariant mass \mfl distribution is used as the discriminating variable
for the \llll final state. It is extracted from simulation for signal events
and for most background components (\ttv, \vvv, \llmumu\ and  heavy-flavour hadron component of \llee),
except for the light-flavour jets and photon conversions in the case  of \llee\ background,
which are taken from the control region as described in Section~\ref{sec:bkg4l}.
 
To obtain statistical interpretations for each mass hypothesis, the \mfl distribution for signal
is parameterised as a function of the mass hypothesis \mH.
In the case of a narrow resonance, the width in \mfl is determined
by the detector resolution, which is modelled by the sum of a Crystal Ball
($\mathcal{C}$) function~\cite{CrystalBall1,CrystalBall2} and a Gaussian ($\mathcal{G}$) function:
$$ P_{s} (\mfl) = f_{\mathcal{C}} \times \mathcal{C}(\mfl; \mu, \sigma_{\mathcal{C}}, \alpha_{\mathcal{C}}, n_{\mathcal{C}}) + (1 - f_{\mathcal{C}}) \times \mathcal{G}(\mfl; \mu, \sigma_{\mathcal{G}}). $$
The Crystal Ball and Gaussian functions share the same peak value of \mfl ($\mu$),
but have different resolution parameters, $\sigma_{\mathcal{C}}$ and $\sigma_{\mathcal{G}}$.
The $\alpha_{\mathcal{C}}$ and $n_{\mathcal{C}}$ parameters control the shape
and position of the non-Gaussian tail, and the parameter $f_{\mathcal{C}}$
ensures the relative normalisation of the two probability density functions.
To improve the stability of the parameterisation in the full mass range considered,
the parameter $n_{\mathcal{C}}$ is set to a fixed value.
The bias in the extraction of signal yields introduced by using the analytic function is below 2\%
and treated as a systematic uncertainty of the signal parameterisation.
The function parameters are determined separately for each final state using the simulated events for each generated mass \mres,
and then fitted with a polynomial in \mres to interpolate between the generated mass points.
The order of the polynomial is determined by first fitting with a third-order polynomial
and then decreasing its order until the $\chi^2$ is three times larger than the number of degrees of freedom.
The use of this parameterisation for the function parameters introduces a bias
in the signal yield and \mres extraction of about 1\%.
The extra bias is included in the systematic uncertainties of the signal acceptance.

In the case of the LWA and the graviton model, a parton-level lineshape of \mfl is derived
from a theoretical calculation and multiplied by the signal acceptance obtained from the simulated events;
it is then convolved with the detector resolution,
using the same functions as those for modelling the narrow resonance.
The parton-level lineshape of \mfl is taken from Ref.~\cite{Goria:2011wa} for the LWA,
and from Ref.~\cite{Bijnens:2001gh} for the graviton model.
 
For the \ZZ continuum background, the \mfl distribution is
parameterised by an empirical function for both the quark- and gluon-initiated processes
in order to reduce the statistical uncertainties stemming from the limited number of simulated events.
The empirical function is described by the following:
\begin{equation*}
f_{qqZZ/ggZZ}(\mfl) = C_0 \times H(m_{0}-\mfl) \times f_1(\mfl) + H(\mfl-m_{0}) \times f_{2} (m_{4\ell}),
\end{equation*}
where,
\begin{equation*}
\begin{split}
f_1(\mfl) &= \Big( \frac{\mfl - a_4}{a_3} \Big)^{a_1 - 1}\Big(1 + \frac{\mfl - a_4}{a_3}\Big)^{-a_1-a_2}, \\
f_2(\mfl) &= \exp\Big[ b_0\Big(\frac{\mfl - b_4}{b_3} \Big)^{b_1 - 1} \Big(1 + \frac{\mfl - b_4}{b_3}\Big)^{-b_1-b_2}\Big], \\
C_0 &= \frac{f_{2} (m_0)}{f_{1} (m_0)}.
\end{split}
\end{equation*}
 
The function's first part, $f_1$, covers the low-mass part of the spectrum until the
\ZZ threshold around $2\cdot m_Z$, and the second part, $f_2$, describes the high-mass tail.
The transition between low- and high-mass parts is modelled with the Heaviside
step function $H(x)$ around $m_0 = 260~\gev{}$ for \qqZZ and around $350~\gev{}$ for \ggZZ.
The continuity of the function around $m_0$ is ensured by the normalisation factor $C_0$ that is applied
to the low-mass part. Finally, $a_i$ and $b_i$ are shape parameters which
are obtained by fitting the \mfl distribution in simulation for each category.
A large number of \mfl distributions are calculated from the analytic function with variations
of the $a_i$ and $b_i$ values sampled from a multivariate Gaussian distribution
that is constructed from their covariance matrix.
The uncertainty in the \mfl distribution is determined by calculating a central interval that captures 68\% of the variations,
and is treated as a nuisance parameter in the likelihood fit, namely a \ZZ parameterisation uncertainty.
The \ZZ parameterisation uncertainty is one of the leading systematic uncertainties for a low-mass signal,
as shown in Table~\ref{tab:ranking}.
 
\subsubsection*{Interference modelling}
The gluon-initiated production of a heavy scalar \BigH, the SM Higgs \SMH and the \ggZZ continuum background
all share the same initial and final state, and thus lead to interference terms in
the total amplitude.  Theoretical calculations described in~Ref.~\cite{Sig-bkg_inter} have shown that the
effect of interference could modify the integrated cross section by up to $\mathcal{O}$(10\%),
and this effect is enhanced as the width of the heavy scalar increases. Therefore, a search for a heavy
scalar Higgs boson in the LWA case must properly account for two interference effects: the
interference between the heavy scalar and the SM Higgs boson (denoted by \IntHh) and between the
heavy scalar and the \ggZZ continuum (denoted by \IntHB).
However, because the width of the KK excitation resonance is relatively small,
the interference effect is assumed to be negligible in the graviton interpretation for both final states.
 
If the \BigH and \SMH bosons have similar properties, they have the same  production and decay amplitudes and therefore the only difference between the signal and interference terms in the production cross section  comes from the propagator. Hence, the acceptance and resolution of the signal and interference terms are expected to be the same.
The \IntHh interference is obtained by reweighting the particle-level lineshape of generated signal events using the following formula:
 
\begin{equation*}
w(\mfl) = \frac{2 \cdot Re \left[\frac{1}{s-s_H} \cdot  \frac{1}{(s-s_h)^*} \right ] }{\frac{1}{\left | s - s_H \right |^2}},
\end{equation*}
where $1/\left(s-s_{H (h)}\right)$ is the propagator for a scalar (\BigH or \SMH).
The particle-level lineshape is then convolved  with the detector resolution function, and the signal and interference acceptances are assumed to be the same.
 
In order to extract the \IntHB interference contribution,
signal-only and background-only samples are subtracted from the generated \textit{SBI} samples.
The extracted particle-level \mfl distribution for the \IntHB interference term
is then convolved with the detector resolution.
 
 
\section{Analysis of \llvv~final state}
\label{sec:hllvvanal}
\subsection{Event selection and categorisation}
 
The \llvv~final state consists of a pair of high-\pt isolated leptons (electrons or muons) and large \met{},
and is subject to larger background contamination than the \llll~channel.
Candidate events are recorded with a combination of multiple single-lepton triggers,
which gives a high efficiency of about 98\% for typical signal processes in the signal region defined in the following.
 
Candidate events are preselected by requiring exactly two electrons or muons with opposite charges and $\pT>20~\gev{}$,
where the electrons (muons) must have $|\eta| < $ 2.47 (2.5).
The leading lepton is further required to have $\pT>30~\gev{}$,
well above the threshold of the single-lepton triggers.
The selected electrons or muons must have a longitudinal impact parameter satisfying $|z_0\sin(\theta)| < 0.5$~mm. The lepton candidates are required to satisfy the same isolation criteria and the same requirement on the transverse impact-parameter significance as used in the \llll~channel (see Section~\ref{sec:commonsel4l}),
which leads to an efficiency above 98\% for typical prompt leptons with $\pt > 30~\gev{}$.
To suppress the $WZ$ background, events containing any additional
lepton satisfying the `loose' identification requirement with $\pt > 7~\gev{}$,
in addition to the other requirements, are rejected.
Requiring the dilepton invariant mass ($m_{\ell\ell}$) to be in the range between 76 and $106~\gev{}$ largely reduces the contamination from the \nonresll{} background, originating from
$t\bar{t}$, $Wt$, $WW$, and $Z\rightarrow\tau\tau$ production.
The data sample after the preselection is dominated by the
\Zjets{} and \nonresll{} processes.
To suppress these backgrounds, a further selection based on \met{} and event topology is applied.
 
Candidate events are required to have $\met>120~\gev{}$, which suppresses the \Zjets{} contamination by several orders of magnitude.
The number of residual \Zjets{} events, which have large fake \met, is further reduced by requiring \metsig~$> 10$,
where \metsig~is the statistical significance of the \met~value against the null hypothesis of zero-\met~\cite{ATLAS-CONF-2018-038}.
Additional selection criteria based on angular variables are imposed to further reject the \Zjets~and \nonresll~background events.
The selection on angular variables is motivated by the desired detector signature,
where the \vecmet is back-to-back with the transverse momentum of the dilepton system.
The azimuthal angle difference between the dilepton system and \vecmet, \dphizmet, must be larger than 2.5 radians, and the selected leptons must be close to each other, with the distance \drll~$=\sqrt{(\Delta\phi_{\ell\ell})^2 + (\Delta\eta_{\ell\ell})^2}<$ 1.8.
Furthermore, the azimuthal angle difference between any of the selected jets with $\pt > 100~\gev{}$ and \vecmet must be larger than 0.4 radians.
As a consequence of all the requirements,
the \Zjets{} process only constitutes a small fraction of the total background (about 4\%) after the full selection.
Finally, events containing one or more $b$-jets are vetoed to further suppress the \ttbar~and $Wt$ backgrounds.
 
The signal region for the VBF production mode (VBF-enriched signal region)
is defined for candidate events containing at least two selected jets with $\pt > 30~\gev{}$,
where the two leading jets must have $\mjj > 550~\gev{}$ and $\detajj > 4.4$.
The remaining events, failing the requirements for the VBF-enriched signal region,
are categorised for the ggF-enriched signal region.
The signal acceptance  in the ggF-enriched signal region for signal events containing a heavy spin-0 resonance from ggF production is
about 30\% at $\mres = 400~\gev{}$ and up to 50\% at $\mres = 1.4~\tev{}$. For VBF signal events the signal acceptance in the VBF-enriched signal region is generally lower, ranging from 3\% at $m_H = 400~\gev{}$ to 20\% at $m_H = 1.6~\TeV{}$.

\subsection{Background estimation}
\label{sec:sigbkgllvv}
 
In the ggF-enriched signal region, the major backgrounds originate from the $ZZ$ and $WZ$ processes,
which account for 60\% and 30\% of the total background contribution, respectively.
The \nonresll~background yields a relative contribution of about 5\% to the total background, while the largely suppressed \Zjets{} background only constitutes a small fraction (4\%).
Finally, the remaining contributions from other processes ($VVV$ and $t\bar{t}V$), amount in total to less than 1\% of the total background.
A similar composition of background processes is found in the VBF-enriched signal region, where the total background yield is expected to be smaller than 1\% of that in the ggF-enriched signal region, due to the event selection for the VBF phase space. The various background estimates and their uncertainties are described below.
 
The main background contribution from $ZZ$ production is estimated using a semi-data-driven method.
Similarly to the \llll~analysis, the predicted $ZZ$ yield is scaled by a floating normalisation factor,
which is determined in the statistical fit to the signal-region data (see Section~\ref{subsec:stats-sys}).
The introduction of the data-driven normalisation factor helps constrain the total uncertainty in the $ZZ$ yield,
while the theoretical and experimental uncertainties in the transverse mass distribution are evaluated from simulation.
 
To estimate the background from $WZ$ production in the ggF-enriched signal region,
a control region enriched in $WZ$ events,
with a purity of over 90\%, is defined using the preselection criteria,
except that a third lepton with $\pt>20~\gev{}$ is required.
Several further selections such as \metsig~$> 3$, a $b$-jets veto, and $m_{\mathrm{T}}^{W}$~$>60~\gev{}$,
where $m_{\mathrm{T}}^{W}$ is constructed from the third lepton's transverse momentum and
the \vecmet{} vector,\footnote{$m_{\mathrm{T}}^{W} = \sqrt{2 \pT^{\ell} \met[1-\cos\Delta\phi({\vec{p}_{\mathrm{T}}^{\ell}},
\vec{E}_{\mathrm{T}}^{\mathrm{miss}})]}$} are applied to suppress non-$WZ$ contributions.
A normalisation factor is calculated in the control region as the number of observed events in data,
after subtracting the non-$WZ$ contributions estimated from simulation, divided by the predicted $WZ$ yield.
The factor is found to be 1.05 with a total uncertainty of 5\%,
which is consistent with a recent $WZ$ measurement~\cite{STDM-2018-03}
performed within a broader fiducial phase space.
The statistical uncertainty of the data in the control region leads to a 0.8\% uncertainty in the $WZ$ estimate in the signal region.
The main systematic uncertainty is evaluated for the ratio of the $WZ$ predictions in the signal and control regions,
and covers the experimental uncertainties and the theoretical ones related to the PDFs and the QCD scales.
The uncertainty related to the subtraction of the non-$WZ$ contribution in the control region is
estimated by applying cross-section uncertainties for all the relevant processes and is found to be negligible.
An additional uncertainty is assigned to the $WZ$ prediction in the signal region,
to account for the efficiency mismodelling of vetoing a third lepton in $WZ\rightarrow\ell\ell\ell\nu$ events.
The total uncertainty in the $WZ$ estimate for the ggF-enriched signal region is about 5\%.
A similar method is adopted to estimate the $WZ$ contribution in the VBF-enriched signal region,
except that the control region additionally selects two jets with $\pt>30~\gev{}$.
The normalisation factor is found to be 0.83 with an uncertainty of 0.27, which is compatible with
the results presented in Ref.~\cite{STDM-2017-23}.
The total uncertainty in the $WZ$ estimate for the VBF-enriched signal region is about 30\%.
The kinematic distributions are estimated from simulation.
 
To estimate the \nonresll~background,
a control region dominated by the \nonresll~processes (with a purity of about 95\%) is defined with all the event selection criteria
except that the final state is required to contain an opposite-sign $e\mu$ pair.
The \nonresll~contribution in the $ee$ ($\mu\mu$) channel is calculated as one half of the observed data yield after subtracting the contribution from the other background processes in the control region, and then corrected for the difference in the lepton reconstruction and identification efficiencies between selecting an $e\mu$ pair and an $ee$ ($\mu\mu$) pair.
The lepton efficiency correction is derived as the square root of the ratio of the numbers of $\mu\mu$ and $ee$ events in data after the preselection.
The choice of deriving the correction after preselection minimises the resulting statistical uncertainty.
The total uncertainty in the \nonresll~estimate in the ggF-enriched signal region is about 9\%,
including the statistical uncertainty of the data in the control region
and the method bias estimated from simulation.
The estimation of the \nonresll~background in the VBF-enriched signal region relies on a similar methodology,
except that the control region is defined with a jet selection that is looser than in the signal region.
The \nonresll~estimate obtained with the looser selection is then scaled by a simulation-based transfer factor to derive the final estimate in the VBF-enriched signal region. The transfer factor is subject to  experimental and theoretical uncertainties, and the relative uncertainty in the final estimate in the VBF-enriched signal region is 70\%.
The kinematic distributions for the \nonresll~background in the signal region are predicted with simulation,
and the assigned systematic uncertainty covers the experimental uncertainty in the simulated shape as well as the difference between data and simulation in the control region.
 
The \Zjets{} background contribution is estimated from simulation and
scaled by a normalisation factor derived in a control region enriched in
\Zjets~events. The control region is defined with all event selection criteria except that \metsig~
must be less than 9 and no requirements on the azimuthal angle difference between jets with
$\pt > 100~\gev{}$ and \vecmet are made.
The normalisation factor is found to be close to one.
Apart from the statistical uncertainty in the control sample,
the experimental and theoretical uncertainties
are evaluated for the ratio of the number of simulated events in the signal region to that in the control region.
The total uncertainty in the \Zjets~estimate is about 40\%.
The kinematic distributions for the \Zjets~background are modelled with simulation.
Finally, backgrounds from the $VVV$ and $t\bar{t}V$ processes,
which contribute less than 1\% of the total background, are estimated from simulation.

\subsection{Signal and background modelling}
\label{sec:modelingllvv}
The modelling of the transverse mass \mt distribution for signal and background
is based on templates derived from fully simulated events and afterwards used to fit the data.
In the case of a narrow resonance, simulated events generated for fixed mass hypotheses
as described in Section~\ref{sec:samples} are used as the inputs in the
moment-morphing technique~\cite{Baak:2014fta} to obtain the \mt distribution for
any other mass hypothesis.
 
The extraction of the interference terms for the LWA case is performed in the same way as in the \llll final state, as described in Section~\ref{sec:modeling4l}.
In the case of the \llvv final state a correction factor, extracted as a function of $m_{ZZ}$, is used
to reweight the interference distributions obtained at particle level to account for reconstruction
effects. The final expected LWA \mt distribution is obtained from the combination of the interference
distributions with simulated \mt distributions, which are interpolated between the simulated mass
points with a weighting technique using the Higgs propagator, a method similar to that used for the interference.
 
\section{Systematic uncertainties}
\label{sec:syst}
The systematic uncertainties can be categorised into experimental and theoretical uncertainties.
The first category includes the uncertainties resulting from the integrated luminosity,
the trigger efficiencies, the momentum scale and resolution of tracks, the reconstruction  and identification of leptons and jets,
and their energy scale and resolution calibrations.
Systematic uncertainties associated with data-driven methods are also in this category, but described
in their corresponding sections:
Section~\ref{sec:bkg4l} for \llll final state and Section~\ref{sec:sigbkgllvv} for \llvv final state.
The second category includes the uncertainties in the theoretical descriptions of the signal
and background simulations.
 
These systematic uncertainties evaluated separately for signal and background
in each category affect signal acceptances and
background yields as well as the probability density distributions of the discriminating variables.
They are provided as the inputs for the statistical interpretations
described in Section~\ref{sec:inter}, in which
the impact of these uncertainties on the expected signal yields are also presented.
 
\subsection{Experimental uncertainties}
The uncertainty in the combined 2015--2018 integrated luminosity is \lumiErr~\cite{ATLAS-CONF-2019-021},
obtained using the LUCID-2 detector~\cite{LUCID2} for the primary luminosity measurements.
 
The lepton identification and reconstruction efficiency and energy/momentum scale and resolution are derived
from data using $J/\psi\rightarrow\ell\ell$ and $Z\rightarrow\ell\ell$ decay events.
The uncertainties in the reconstruction performance are computed following the method described
in Ref.~\cite{PERF-2015-10} for muons and Ref.~\cite{PERF-2017-01} for electrons.
In general, their impact on the signal and background yields is less than 1\% in the \llvv final state,
and up to 1.5\% in the \llll final state.
In addition, the lepton isolation uncertainty is estimated to be less than 1\% in both final states.
 
The uncertainties in the jet energy scale and resolution have several sources, including
uncertainties in the absolute and relative \textit{in situ} calibration, the correction for \pu,
the flavour composition and response~\cite{PERF-2016-04}.
Each source is treated as an independent component.
They vary from 4.5\% for jets with transverse momentum $\pt = 20~\gev{}$,
decreasing to 1\% for jets with $\pt = 100$--$1500~\gev{}$ and increasing again to 3\% for jets with higher \pt.
They are the dominant uncertainties in the VBF-enriched categories for ggF signal production and SM $ZZ$ production
in both final states.
 
Uncertainties in the lepton and jet energy scales are propagated to the uncertainty in the \met~\cite{PERF-2016-07}.
Additionally, the uncertainties from the momentum scale and resolution of the tracks that are not
associated with any identified lepton or jet contribute 8\% and 3\%, respectively, to the uncertainty in the \met value.
 
The efficiency of the lepton triggers in events with reconstructed leptons is nearly 100\%,
and hence the related uncertainties are negligible.
The uncertainties associated with the \pu reweighting are also taken into account;
their impact on the signal and background yields is about 1\% for both final states.
 
These experimental uncertainties are common to the two final states; therefore, they are
fully correlated between the two final states.

\subsection{Theoretical uncertainties}
For the simulation-based estimates, the theoretical uncertainties stemming from
parton distribution functions (PDFs), missing higher-order QCD corrections, and parton showering
are considered.
 
The PDF uncertainty is evaluated by taking the envelope of variations among alternative PDF choices and
the estimate from its internal PDF error sets, following the PDF4LHC recommendation~\cite{Butterworth:2015oua}.
The missing higher-order QCD corrections are estimated by halving or doubling the factorisation and renormalisation
scales independently, among which the largest effect is taken as the systematic uncertainty.
The parton-showering uncertainty is assessed by varying the \PYTHIA~configurations,
such as the parameter values of the AZNLO tune,
the multi-parton models and the final-state radiation models.
 
For different signal hypotheses, the impact of these theoretical uncertainties on the signal
acceptance and the spectrum of the discriminating variables is evaluated.
In total, the theoretical uncertainty in the signal acceptance varies
from less than 1\% in the low mass region
to 12\% in the high mass region of the \llvv final state,
and from less than 1\% in the low mass region to up to 20\%
in the high mass region of the \llll final state.
 
For the continuum $ZZ$ background, a common floating normalisation factor is
introduced to scale the number of events for the \qqZZ and \ggZZ processes, while
the relative yields of the two processes are estimated from the simulations.
Therefore, in addition to the spectrum of the discriminating variables in the $ZZ$ background,
the theoretical uncertainties are also propagated to the simulation-based estimation of the relative yields.
Moreover, the uncertainty associated with
the NLO EW corrections, calculated in Refs.~\cite{Biedermann:2016yvs,Biedermann:2016lvg,Kallweit:2017khh},
are also taken into account, affecting the discriminating variables by less
than 1\% in the low mass region and up to 10\% in the high mass region for both final states.

Because the \llll and \llvv searches are sensitive to
different energy scales, these theoretical uncertainties are assumed to be completely uncorrelated between
the two analyses. A fully correlated scenario is also examined and the differences
between the two scenarios in terms of the expected limits on various signal hypotheses are negligible.

 
\section{Results}
\label{sec:results}
The statistical procedure used  to extract the results is described in Section~\ref{subsec:stats-sys}
and the results are presented in Section~\ref{subsec:gr}.
 
\subsection{Statistical procedure and impact of systematic uncertainties}
\label{subsec:stats-sys}
The statistical treatment of the data interpretation follows the procedure
for the Higgs-boson search combination in 7~\tev{} data~\cite{statscomb, HIGG-2012-17}.
The test statistic used for limit setting is the profile likelihood ratio
$\Lambda(\alpha, \boldsymbol{\theta})$,
which depends on one or more parameters of interest $\alpha$,
additional normalisation factors and extra nuisance parameters $\boldsymbol{\theta}$.
The parameter of interest is the cross section times branching ratio of the heavy resonance decaying into the two final states.
The normalisation factors, which were not used in the previous publication~\cite{HIGG-2016-19},
are introduced separately for each final state
to scale the expected number of the SM \ZZ background events in each category
and are determined by a likelihood fit to the data. This allows the systematic uncertainty to be reduced
by removing both the theoretical and luminosity uncertainties contributing to the normalisation uncertainty. In the \llll final state,
three floating normalisation factors are introduced for the \vbfenr, \ggfhigh and \ggflow categories.
They are referred to as \muZZVBF, \muZZggFHi and \muZZggFLow, respectively.
The use of three $ZZ$ normalisation factors for the \llll~final state is
motivated by the different phase spaces defined for the respective signal regions.
Only one floating normalisation factor \muZZlv is introduced in the \llvv final state,
due to the limited size of the data sample and the worse signal-to-background ratio
in the respective VBF-enriched signal region.

The nuisance parameters represent the estimates of the systematic uncertainties
and each of them is constrained by a Gaussian distribution.
For each category of each final state, a discriminating variable is used to further separate signal from background.
The number of signal events is extracted from a simultaneous fit to the discriminating variable, \mfl in the \llll analysis and \mt in the \llvv analysis, in the event categories described in Sections~\ref{sec:h4lanal} and \ref{sec:hllvvanal}.
 
The impact of a systematic uncertainty on the result depends on the production mode
and the mass hypothesis.
For the ggF production mode, at lower masses the \ZZ parameterisation for the \llll final state
and the systematic uncertainty of the \Zjets\ background for the \llvv final state dominate,
and at higher masses the uncertainties in the NLO EW correction and parton showering become important,
as also seen in VBF production.
For the VBF production mode, the dominant uncertainties come from
the theoretical modelling of the discriminating variables of the \ZZ events in the VBF category.
At lower masses, jet-energy-scale uncertainties are also important.
Table~\ref{tab:ranking} shows the impact of the leading systematic uncertainties
on the predicted signal event yield
when the cross section times branching ratio is set to the expected upper limit (shown in Figure~\ref{fig:NWAlimits}),
for ggF and VBF production modes.
The statistical uncertainty of the data sample dominates in both of the present searches,
and the systematic uncertainties impact the searches to a much lesser extent.

\begin{table}[htbp!]
\caption{
Impact of the leading systematic uncertainties, the data statistical uncertainties
and the total uncertainties on the
predicted signal event yield with the cross section times branching ratio being
set to the expected upper limit,
expressed as a percentage of the signal yield
for the ggF (left) and VBF (right) production modes at $\mres = 300$, 600, 1000, and $1500~\gev$.
}
\label{tab:ranking}
\footnotesize
\centering
 
\begin{tabular}{l S[table-format=2.1] | l S[table-format=2.1]}
\toprule
\multicolumn{2}{c|}{ggF production} &  \multicolumn{2}{c}{VBF production} \\
\multicolumn{2}{l|}{Systematic source    \hfill Impact [\%]}   & \multicolumn{2}{l}{Systematic source \hfill Impact [\%]}\\
\midrule
\multicolumn{4}{c}{$\bigstrut$  $\mres=300~\gev$}\\
\midrule
 
\zzFnc            & 4.5 & \jfc      & 3.0 \\
\abcd             & 2.3 & \qcdvvVBF & 2.8 \\
\shwggf\ (\llll)  & 2.2 & \zzFnc    & 2.3 \\
\emuemu           & 2.0 & \jesInsu  & 1.8 \\
Data stat. uncertainty &  53 & Data stat. uncertainty & 58 \\
Total uncertainty &  55 & Total uncertainty & 60 \\
\midrule
\multicolumn{4}{c}{$\bigstrut$  $\mres=600~\gev$}\\
\midrule
 
\zzEWCorr      & 4.9  & \qcdvv (\llvv) & 7.6 \\
\qcdvvllvv     & 2.5  & \jer           & 5.4 \\
\abcd          & 2.5  & \shwlv         & 3.3 \\
\pdfZZ (\llll) & 2.2  & \zzEWCorr      & 3.0 \\
Data stat. uncertainty &  54 & Data stat. uncertainty & 61 \\
Total uncertainty &  57 & Total uncertainty & 63 \\
\midrule
\multicolumn{4}{c}{$\bigstrut$  $\mres=1000~\gev$}\\
\midrule
 
\zzEWCorr   & 9.3   & \shwlv      & 6.8 \\
\shwlv      & 5.2   & \zzEWCorr   & 4.7 \\
\qcdvvllvv  & 4.8   & \qcdvvllvv  & 2.4 \\
\abcd       & 2.4   & \jfc        & 2.4 \\
Data stat. uncertainty &  57 & Data stat. uncertainty & 58 \\
Total uncertainty &  59 & Total uncertainty & 59 \\
\midrule
\multicolumn{4}{c}{$\bigstrut$  $\mres=1500~\gev$}\\
\midrule
\shwlv      & 9.6   & \shwlv      & 9.0 \\
\zzEWCorr   & 6.8   & \zzEWCorr   & 4.6 \\
\pdfZZlv    & 5.4   & \pdfZZlv    & 3.4 \\
\qcdvvllvv  & 4.6   & \qcdvvllvv  & 2.8 \\
Data stat. uncertainty &  57 & Data stat. uncertainty & 55 \\
Total uncertainty &  59 & Total uncertainty & 57 \\
\bottomrule
 
\end{tabular}
\end{table}
 
\subsection{General results}
\label{subsec:gr}
The total number of observed events is 3275 in the \llll final state ($\mfl > 200~\gev{}$)
and 2794 in the \llvv final state.
The expected background yields are obtained from a simultaneous likelihood fit of the two final states
under the background-only hypothesis.
The fitted normalisation factors for the SM $ZZ$ background are
summarised in Table~\ref{tab:fittedFactors}.
 
\begin{table}
\caption{
The \ZZ normalisation factors together with their total (statistical+systematic)  uncertainties in each category of the two final states,
which scale the number of \ZZ events estimated from the simulations,
obtained from a simultaneous likelihood fit of the two final states
under the background-only hypothesis.
For the \llll final state, the MVA-based categorisation is used.
}
\label{tab:fittedFactors}
\centering
\begingroup
\setlength{\tabcolsep}{10pt}
\renewcommand{\arraystretch}{1.5}
\begin{tabular}{c|
c|
S[table-format=1.2]  @{ $\pm$ } S[table-format=1.2]  }
\hline
Final state & Normalisation factor & \multicolumn{2}{c}{Fitted value} \\ \hline
\multirow{3}{*}{\llll} & \muZZVBF    &  0.9  & 0.3  \\
& \muZZggFHi  &  1.07 & 0.05 \\
& \muZZggFLow &  1.12 & 0.03 \\
\midrule
\llvv                  & \muZZlv     &  1.04 & 0.06 \\ \hline
\end{tabular}
\endgroup
\end{table}
 
\begin{figure}[tbp]
\vspace{-1cm}
 
\subfigure[]{\includegraphics[width=0.32\textwidth]{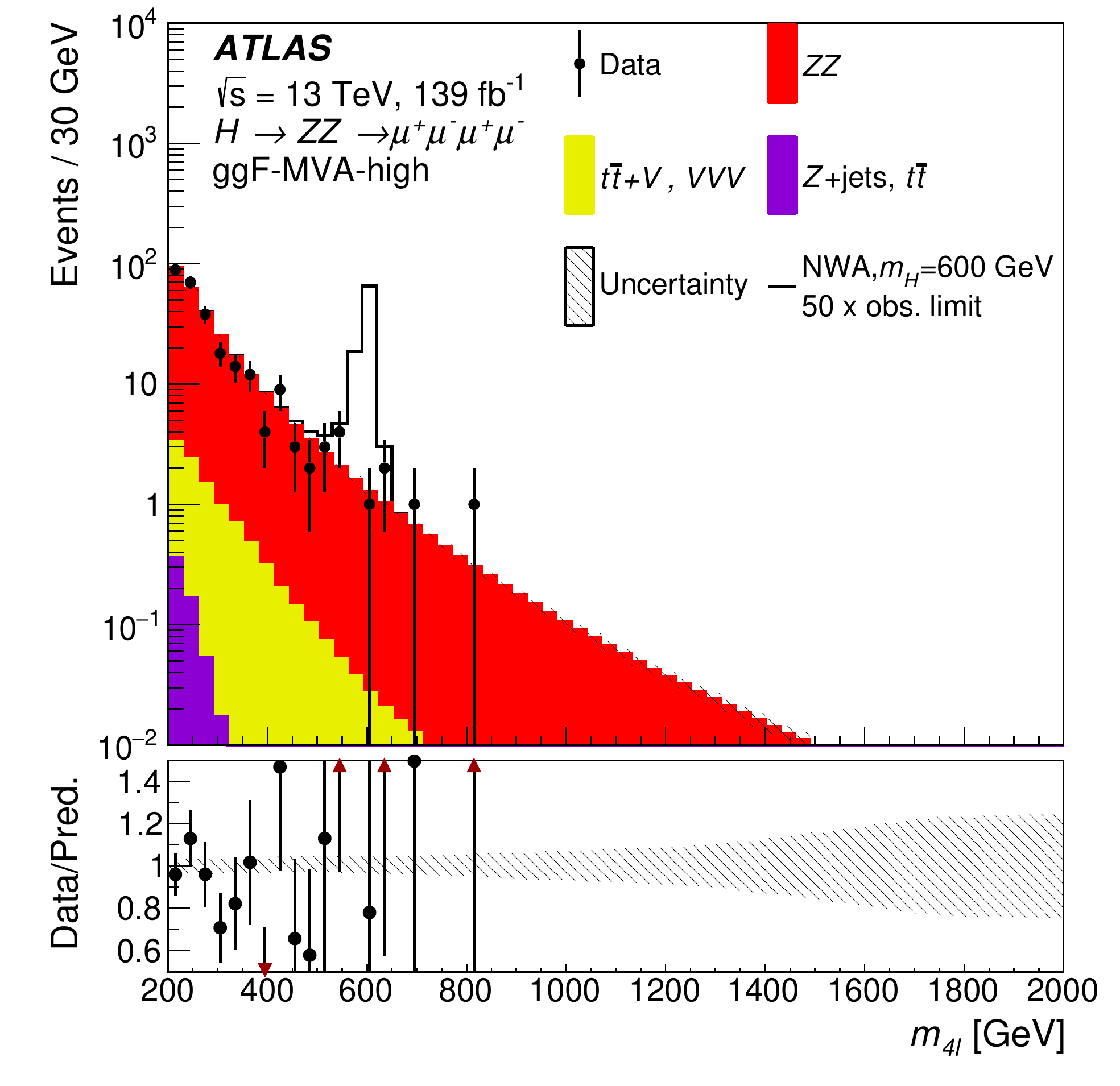}}
\subfigure[]{\includegraphics[width=0.32\textwidth]{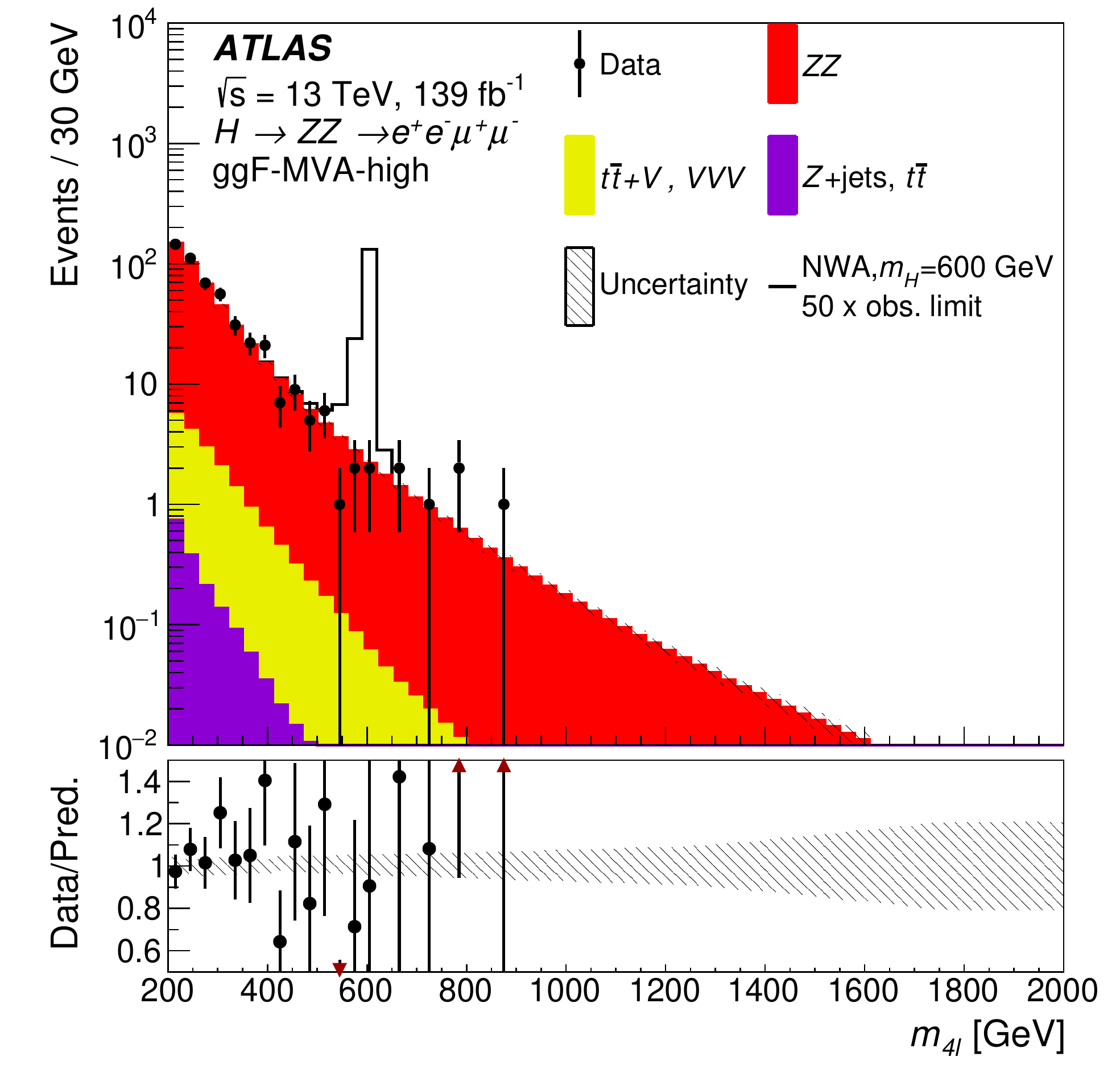}}
\subfigure[]{\includegraphics[width=0.32\textwidth]{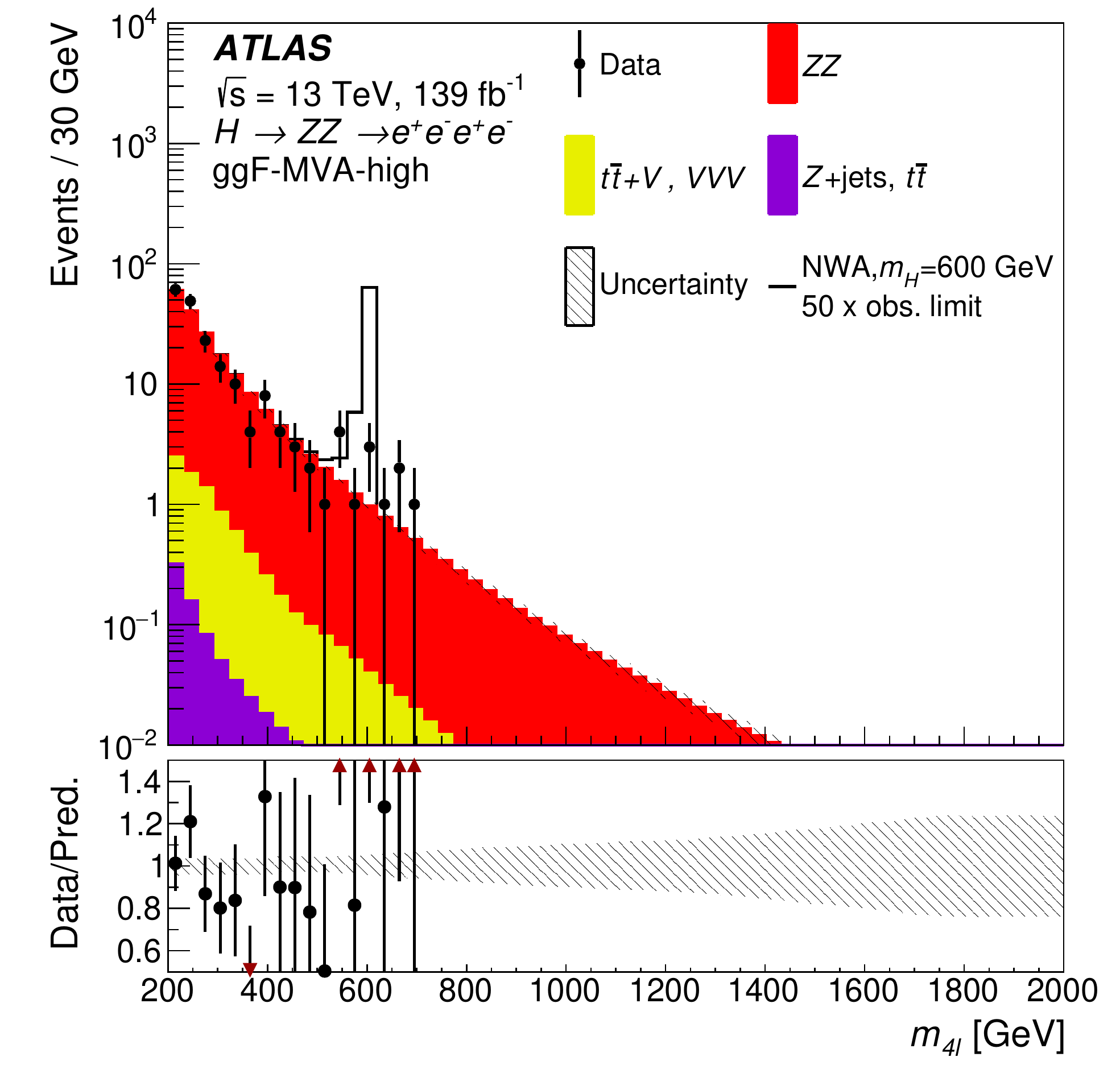}}
\centering
\subfigure[]{\includegraphics[width=0.33\textwidth]{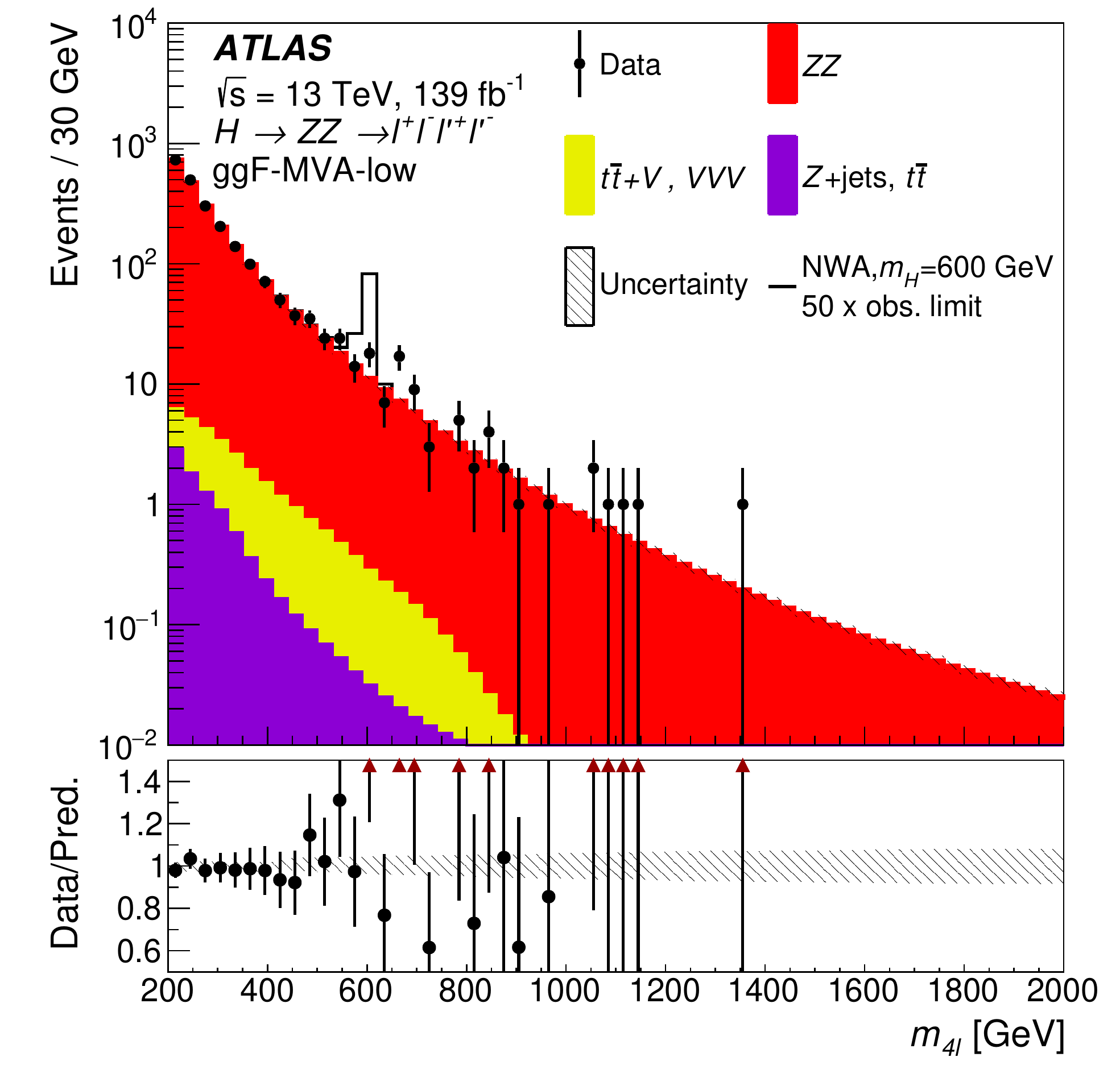}}
\subfigure[]{\includegraphics[width=0.33\textwidth]{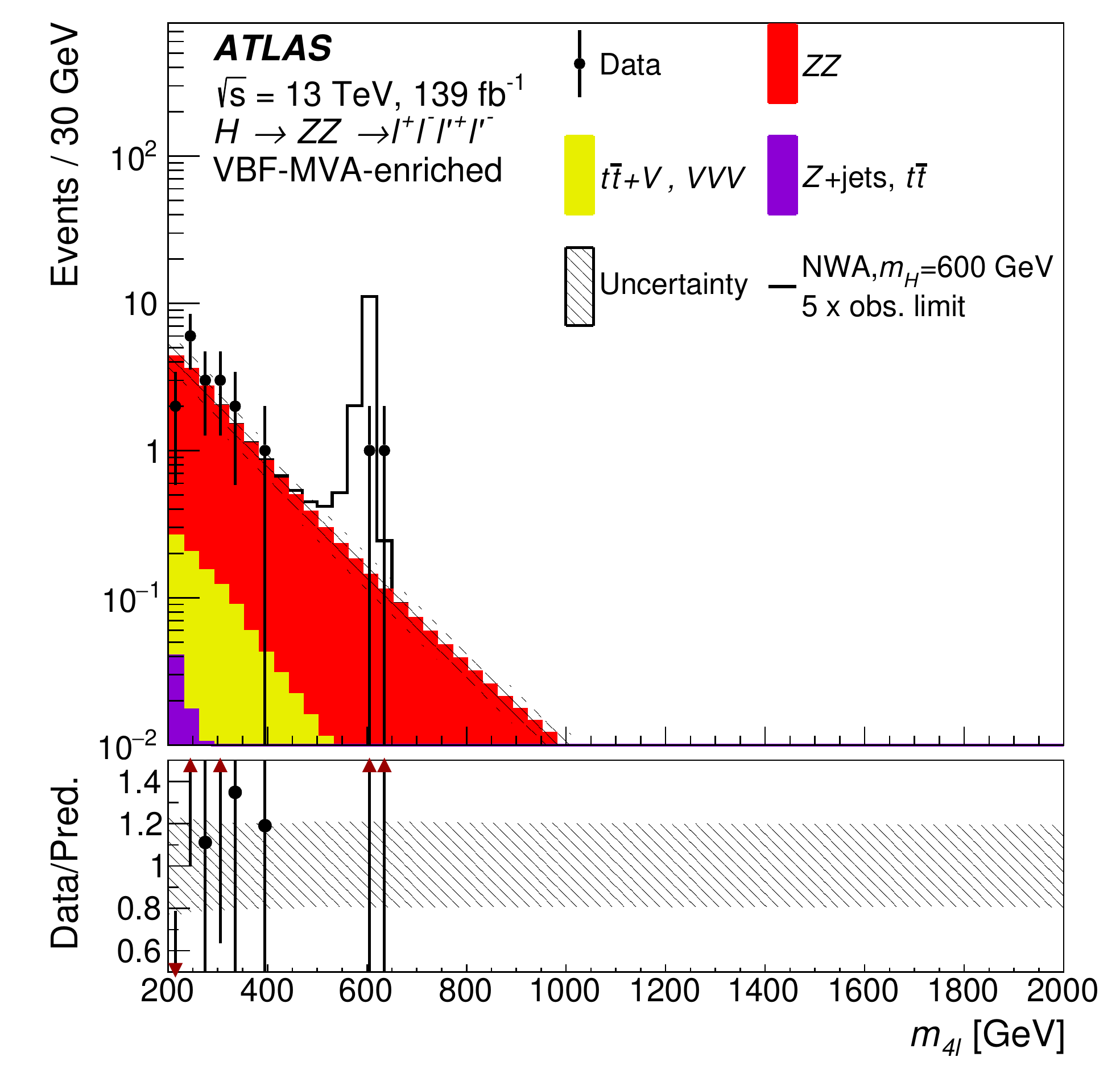}}
 
\caption{Distributions of the four-lepton invariant mass \mfl in the $\llll$ search
for the \ggfhigh categories (\mmmm (a), \eemm (b),
and \eeee (c) final states),
for the \ggflow category (d),
and for the \vbfMVAenr category (e).
The backgrounds are determined from a combined likelihood fit to the data under the background-only
hypothesis. The simulated $\mres = 600~\gev{}$ signal is normalised to
a cross section corresponding to 50 (5) times the observed limit given in Section 9.1.1 for the ggF (VBF) production mode.
The error bars on the data points indicate the statistical uncertainty,
while the systematic uncertainty in the prediction is shown by the \systband\ band.
The lower panels show the ratio of data to prediction. The red arrows indicate data points that are outside the displayed range.
}
\label{fig:m4l_sig}
\end{figure}

\begin{figure}[tbp]
\subfigure[]{\includegraphics[width=0.49\textwidth]{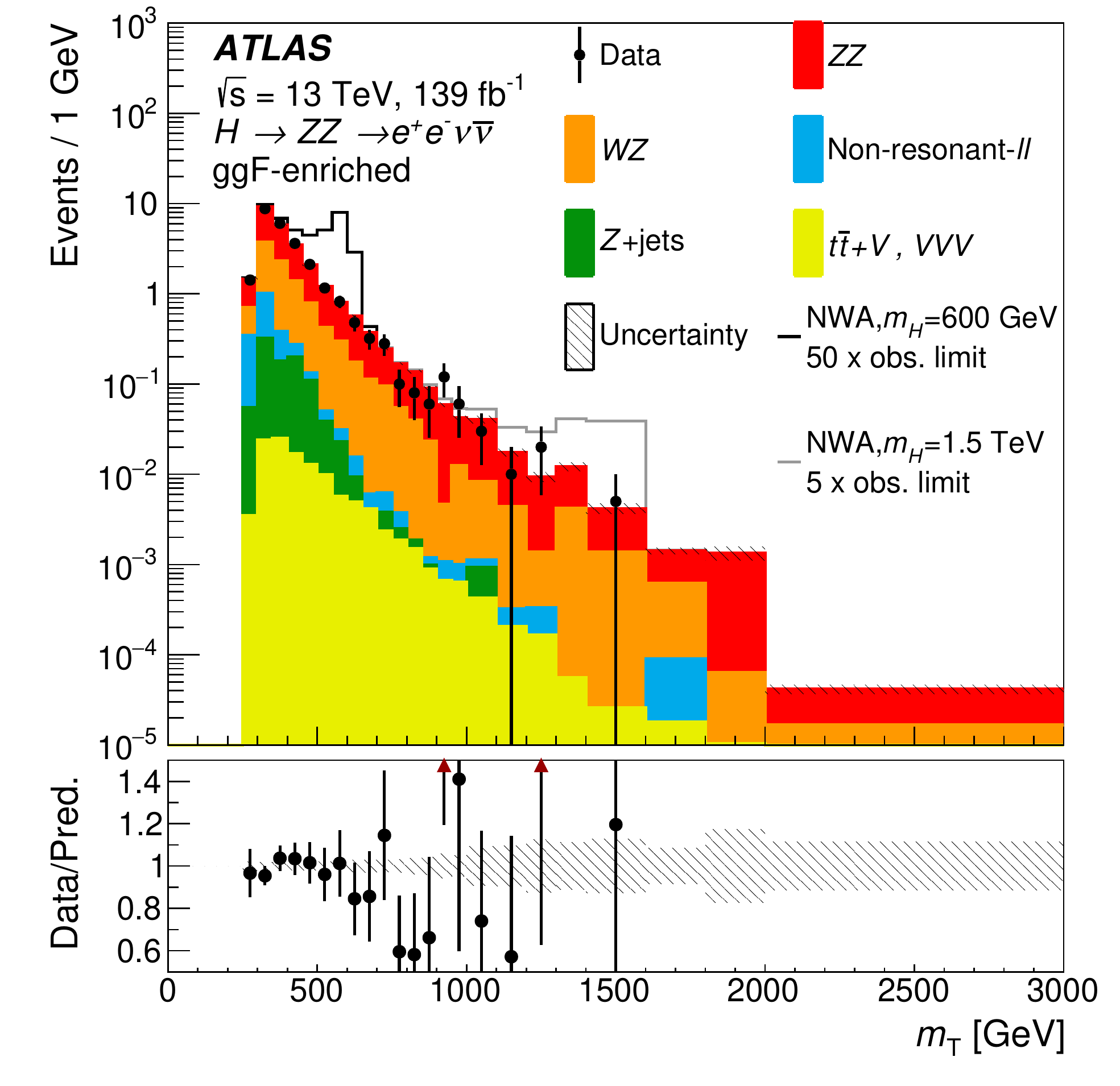}}
\subfigure[]{\includegraphics[width=0.49\textwidth]{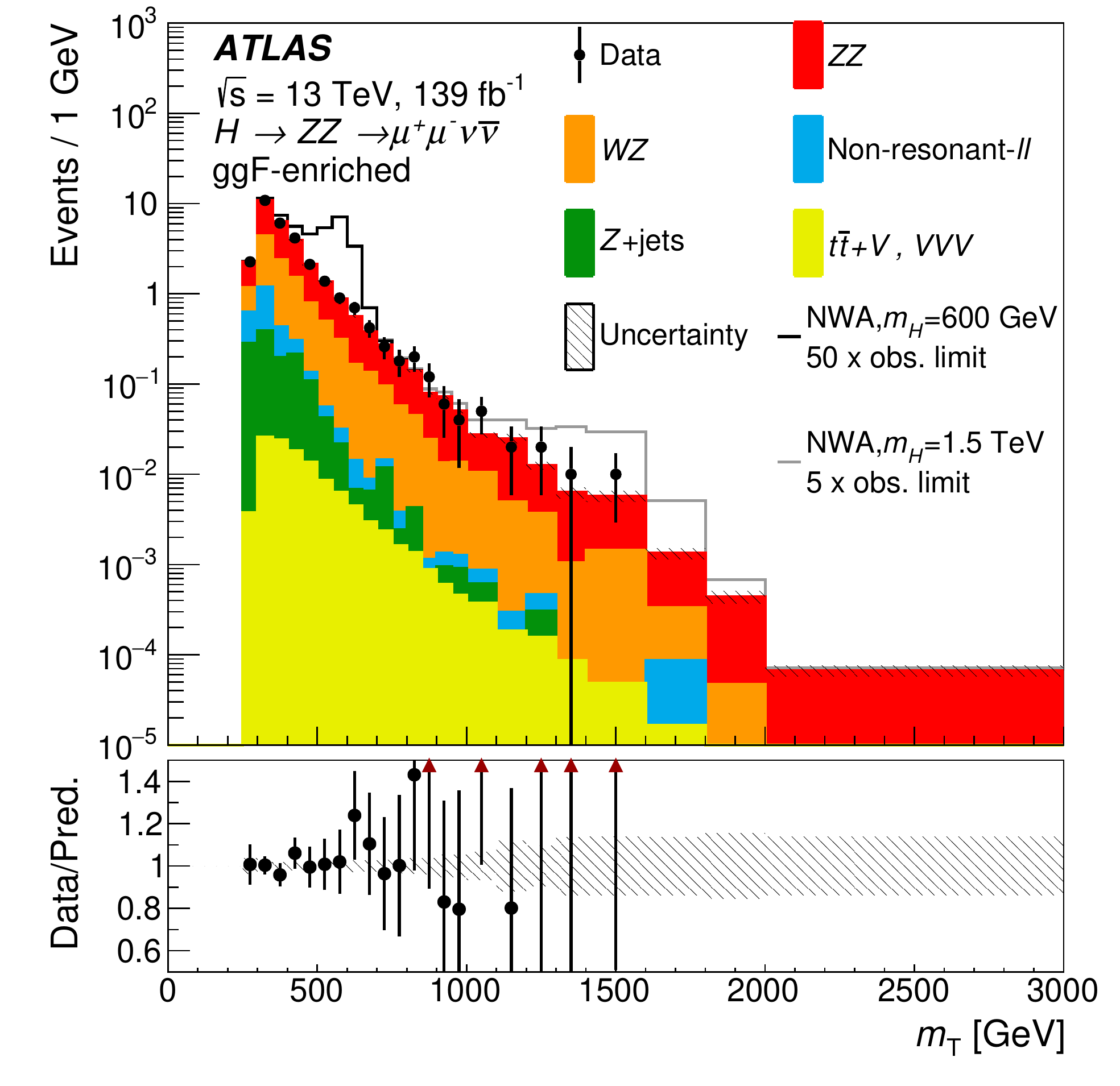}}
\subfigure[]{\includegraphics[width=0.49\textwidth]{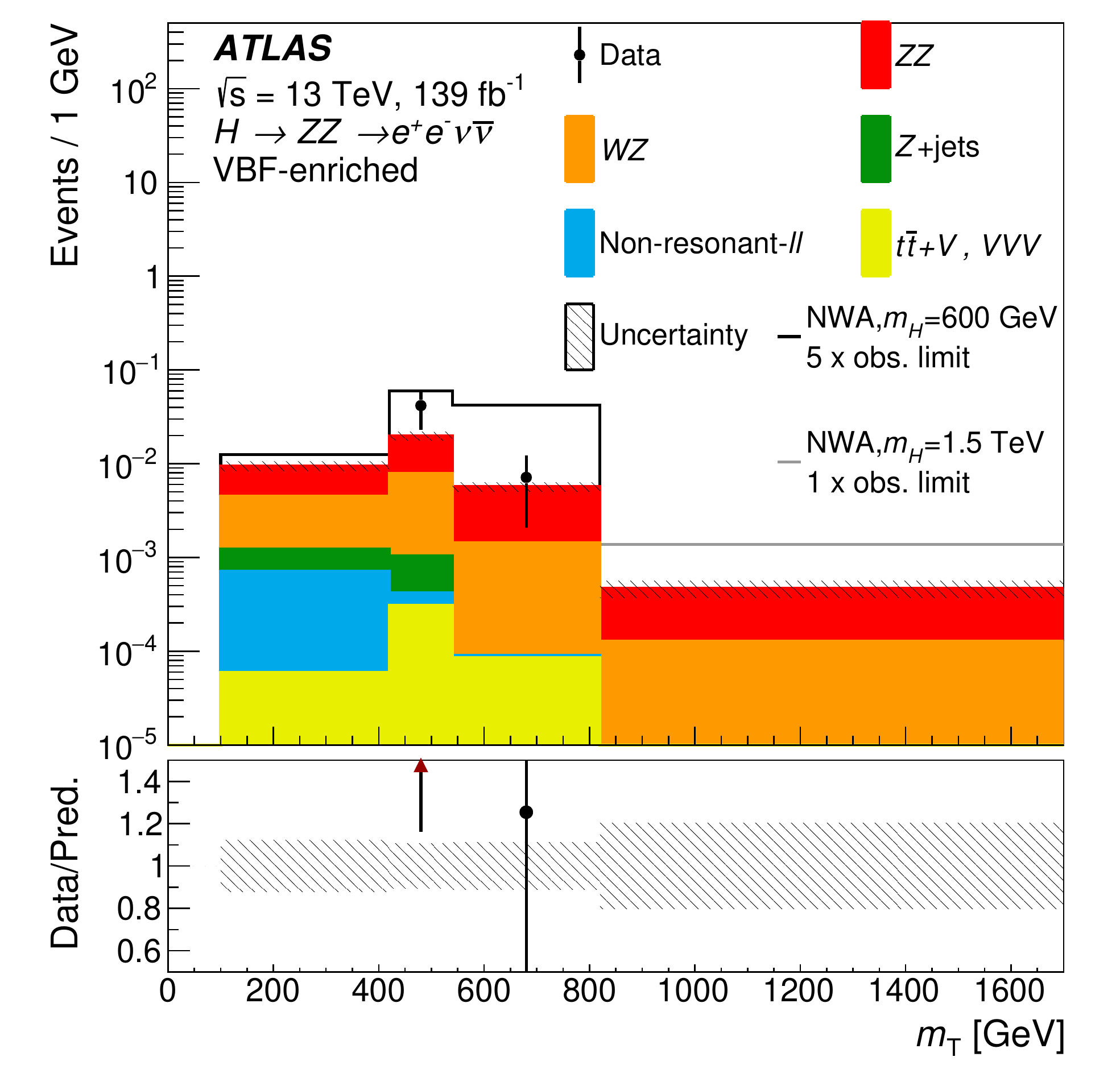}}
\subfigure[]{\includegraphics[width=0.49\textwidth]{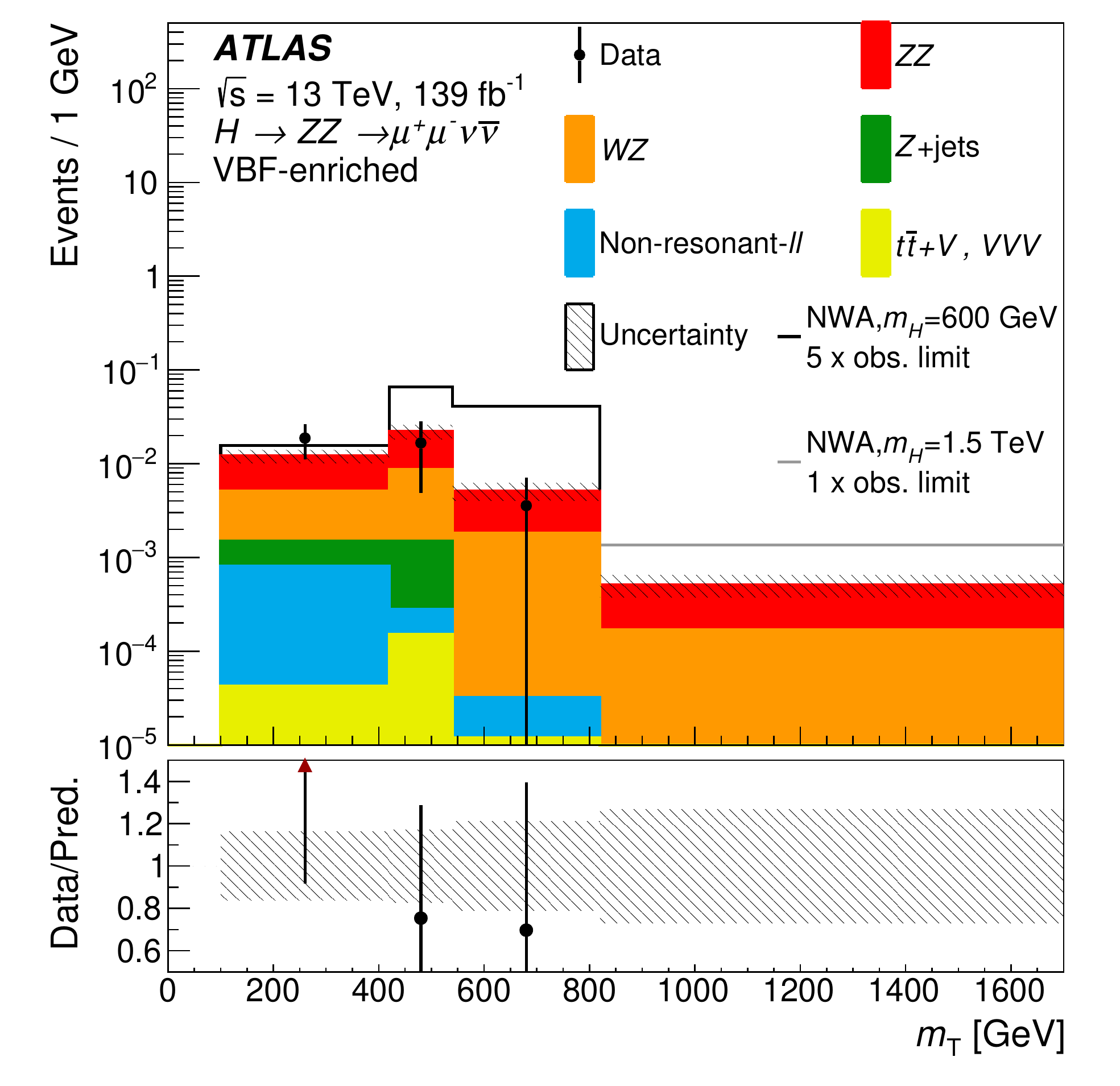}}
\caption{The \mt{} distribution in the $\llvv$ search
for (a),(b) the ggF categories
and (c),(d) the VBF categories.
Events beyond the upper limit of the histogram are included in the last bin of the distribution.
The backgrounds are determined from a combined likelihood fit to data under the background-only
hypothesis.
The simulated $\mres = 600~\gev{}$ ($1.5~\tev{}$) signals are normalised to a cross section corresponding
to 50 (5) times the observed limit given in Section~\ref{sec:spin0nwa}  for the ggF production mode and to 5 (1) times the observed limit for the VBF production mode.
The error bars on the data points indicate the statistical uncertainty and
markers are drawn at the bin centre. The systematic uncertainty in
the prediction is shown by the \systband\ band.
The lower panels show the ratio of data to prediction. The red arrows indicate data points that are outside the displayed range.
}
\label{fig:llvv_sig}
\end{figure}

The number of observed candidate events with mass above $200~\gev{}$ together with the expected background yields for each of the five categories of the \llll analysis as described in Section~\ref{sec:mvasel4l} are presented in Table~\ref{tab:H4lyields}. The \mfl spectrum in each category is shown in Figure~\ref{fig:m4l_sig} for illustration, since the backgrounds are determined from a combined unbinned likelihood fit to the data under the background-only hypothesis.
Table~\ref{tab:ZZllnn_yields_postfit} contains the number of observed events along with the obtained background yields for the \llvv analysis and Figure~\ref{fig:llvv_sig} shows the \mt\ distribution for the electron and muon channels in the ggF-enriched and  VBF-enriched categories.

The two previous excesses around $240~\gev{}$ and $700~\gev{}$ that were observed in the publication~\cite{HIGG-2016-19} using 2015 and 2016 data are not confirmed using the full Run~2 dataset as explained below. The maximum deviation of the data from the background-only hypothesis is evaluated in the context of a NWA signal from the ggF production or from the VBF production separately. For the ggF production, the maximum deviation is for a signal mass hypothesis around 240 GeV, with a local significance of 2.0 standard deviations and a global significance of 0.5 standard deviation. For the VBF production, the maximum deviation is for a signal mass hypothesis around 620 GeV, with a local significance of 2.4 standard deviations and a global significance of 0.9 standard deviation.

\begin{table}[tbp]
\caption{
Expected and observed numbers of events in the \llll final state for $\mfl > 200~\gev{}$,
together with their uncertainties,
for the \vbfMVAenr, \ggfhigh and \ggflow categories.
The expected numbers of events, as well as their uncertainties, are obtained
from a combined likelihood fit to the data under the background-only hypothesis.
The uncertainties of the \ZZ normalisation factors, presented in Table~\ref{tab:fittedFactors},
are also taken into account.
}
\label{tab:H4lyields}
\centering
 
\tiny
\resizebox{\textwidth}{!}{
\begin{tabular}{l
S[table-format=2.2]  @{ $\pm$ } S[table-format=1.2]
S[table-format=3.1]  @{ $\pm$ } S[table-format=2.1]
S[table-format=3.1]  @{ $\pm$ } S[table-format=2.1]
S[table-format=3.1]  @{ $\pm$ } S[table-format=1.1]
S[table-format=4.1]  @{ $\pm$ } S[table-format=2.1]
}
\toprule
\multirow{2}{*}{Process} & \multicolumn{2}{c}{\vbfenr} & \multicolumn{6}{c}{\ggfhigh}  & \multicolumn{2}{c}{\ggflow} \\
& \multicolumn{2}{c}{} & \multicolumn{2}{c}{$4\mu$ channel} & \multicolumn{2}{c}{$2e2\mu$ channel} & \multicolumn{2}{c}{$4e$ channel}  & \multicolumn{2}{c}{} \\
\midrule
 
\qqZZ	&	11	&	4	&	232	&	10	&	389	&	17	&	154	&	7	&	2008	&	47	\\
\ggZZ	&	3	&	2	&	37	&	6	&	64	&	10	&	26	&	4	&	247	&	19	\\
\ZZEW	&	4.1	&	0.4	&	4.5	&	0.2	&	7.5	&	0.4	&	3	&	0.2	&	14.3	&	0.7	\\
\Zjets, \ttbar	&	0.08	&	0.02	&	0.6	&	0.1	&	1.7	&	0.4	&	0.8	&	0.1	&	8.8	&	2.1	\\
\ttv, \vvv	&	0.96	&	0.1	&	9.8	&	0.2	&	17.5	&	0.4	&	7.7	&	0.2	&	21.9	&	0.5	\\
\midrule
Total background        &       19      &       5     &       284     &       12    &       480     &       20    &       192     &       8     &       2300    &       51    \\
 
\midrule
Observed  & \multicolumn{2}{l}{19} & \multicolumn{2}{l}{271} & \multicolumn{2}{l}{493} & \multicolumn{2}{l}{191} & \multicolumn{2}{l}{2301} \\
\bottomrule
\end{tabular}
}
\end{table}
 
\begin{table}[tbp]
\caption{
Expected and observed numbers of events together with their uncertainties in the \llvv final state,
for the ggF- and VBF-enriched categories.
The expected numbers of events, as well as their uncertainties, are obtained
from a likelihood fit to the data under the background-only hypothesis.
The uncertainties of the \ZZ normalisation factors, presented in Table~\ref{tab:fittedFactors},
are also taken into account.
}
\label{tab:ZZllnn_yields_postfit}
\centering
 
\begin{tabular}{l
S[table-format=4.1]  @{ $\pm$ } S[table-format=2.1]
S[table-format=4.1]  @{ $\pm$ } S[table-format=2.1]
S[table-format=1.2]  @{ $\pm$ } S[table-format=1.2]
S[table-format=2.2]  @{ $\pm$ } S[table-format=1.2]
}
\toprule
\multirow{2}{*}{Process} & \multicolumn{4}{c}{\ggfenr}  & \multicolumn{4}{c}{\vbfenr} \\
& \multicolumn{2}{c}{\ee channel} & \multicolumn{2}{c}{\mm channel} & \multicolumn{2}{c}{\ee channel} & \multicolumn{2}{c}{\mm channel} \\
\midrule
 
\qqZZ	&	695	&	39	&	795	&	44	&	2.8	&	0.2	&	3.3	&	0.2	\\
\ggZZ	&	87	&	28	&	97	&	31	&	1	&	0.4	&	1	&	0.4	\\
\ZZEW	&	6.6	&	0.5	&	7	&	0.5	&	0.8	&	0.1	&	0.9	&	0.1	\\
\WZ	&	400	&	13	&	443	&	12	&	2.4	&	0.5	&	3	&	1.3	\\
\Zjets	&	39	&	12	&	56	&	21	&	0.2	&	0.2	&	0.3	&	0.3	\\
Non-resonant-$\ell\ell$	&	66	&	6	&	77	&	7	&	0.2	&	0.2	&	0.3	&	0.2	\\
\ttv, \vvv	&	5.9	&	0.4	&	5.9	&	0.4	&	0.08	&	0.02	&	0.04	&	0.01	\\
\midrule
Total backgrounds	&	1299	&	52	&	1480	&	59	&	7.4	&	0.7	&	8.4	&	1.4	\\

\midrule
Observed                 &\multicolumn{2}{l}{1280} & \multicolumn{2}{l}{1498}  & \multicolumn{2}{l}{7} &  \multicolumn{2}{l}{9} \\
 
\bottomrule
\end{tabular}
\end{table}
 
\clearpage
 
\section{Interpretations}
\label{sec:inter}
Since no significant excess with respect to the background predictions is found, results obtained from the combination of the \llll and \llvv final states are interpreted in terms of exclusion limits for different signal hypotheses as presented below.

\subsection{Spin-0 resonances}
 
\subsubsection{Spin-0 resonances with NWA}
\label{sec:spin0nwa}
\begin{figure}[htbp]
\subfigure[]{\includegraphics[width=0.49\textwidth]{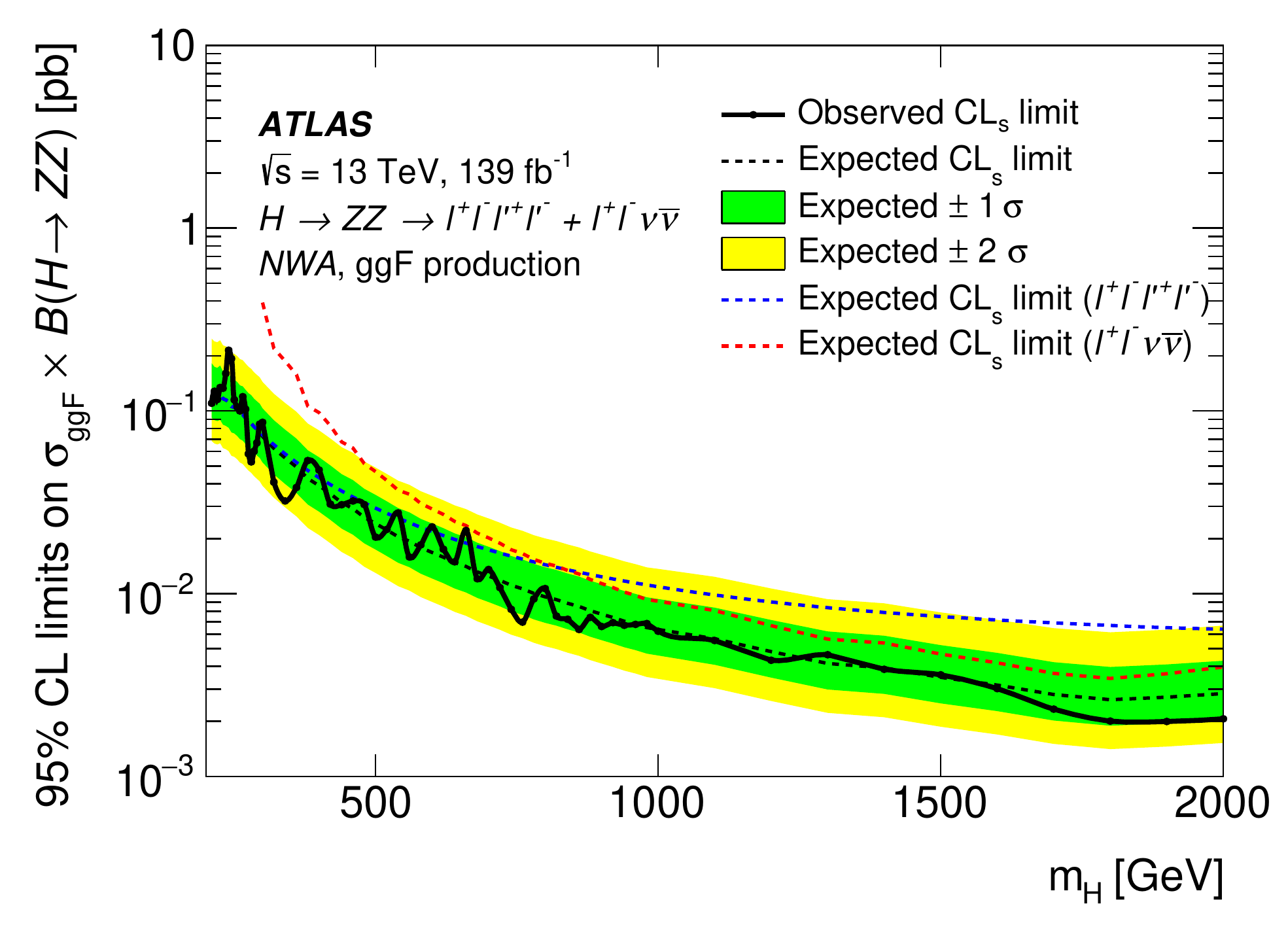}}
\subfigure[]{\includegraphics[width=0.49\textwidth]{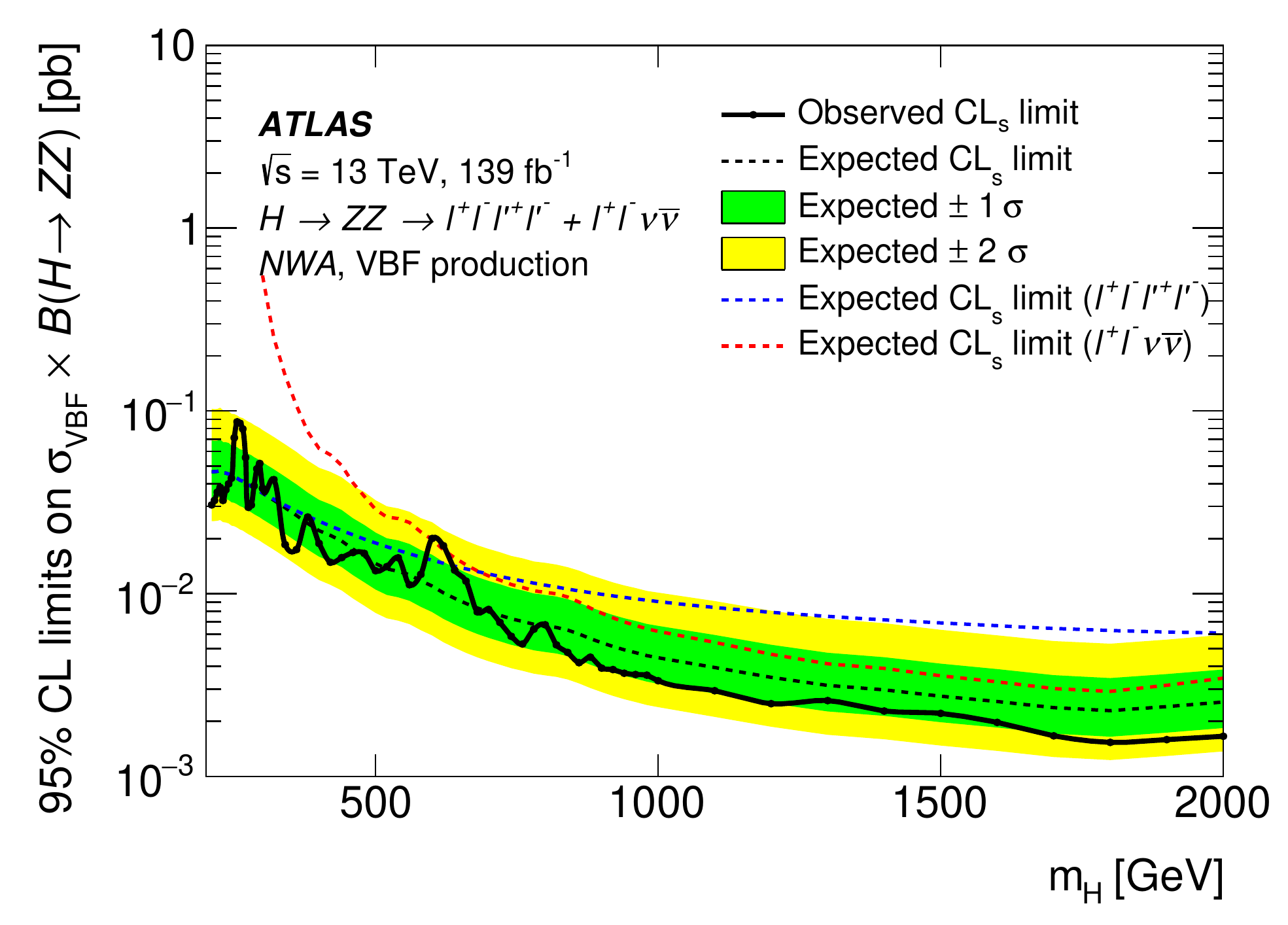}}
\caption{The upper limits at 95\% CL  on the cross section times branching ratio
as a function of the heavy resonance mass $\mres$ for (a)
the ggF production mode ($\xsggF\times\mbr(\htozz)$) and (b)
for the VBF production mode  ($\xsVBF\times\mbr(\htozz)$)  in the case of the NWA.
The black line indicates the observed limit.
The green and yellow  bands represent the $\pm$$1\sigma$ and $\pm$$2\sigma$ uncertainties
in the expected limits. The dashed coloured lines indicate the expected limits obtained
from the individual searches.
}
\label{fig:NWAlimits}
\end{figure}

Upper limits on the cross section times branching ratio ($\sigma \times \mbr (\htozz)$)
for a heavy resonance are obtained from the combination of the two final states, as a function of $\mres$ with the $\mathrm{CL}_{\mathrm{s}}$  procedure~\cite{Read:2002hq} in the asymptotic approximation.
The results were verified to be correct within about 4\% using pseudo-experiments.
It is assumed that an additional heavy scalar would be produced mainly via the ggF and VBF processes but that the ratio of the two production mechanisms might depend on the model considered.
For this reason, fits for the ggF and VBF processes are done separately, and in
each case the other process is allowed to float in the fit as an additional free parameter.
Figure~\ref{fig:NWAlimits} presents the observed and expected limits at 95\% CL on
the $\sigma \times\mbr(\htozz)$ of a narrow scalar resonance for
the ggF (left) and VBF (right) production modes, as well as the expected limits from the $\llll$ and
$\llvv$ searches.
This result is valid for models in which the width is less than 0.5\% of \mres.
When combining the two final states, the 95\% CL upper limits range from 215~fb at $\mres = 240~\gev{}$
to 2.0~fb at $\mres = 1900~\gev{}$ for the ggF production mode
and from 87~fb at $\mres = 255~\gev{}$ to 1.5~fb at $\mres = 1800~\gev{}$
for the VBF production mode.
Compared with the expected limits projected to the luminosity of \lumi\ from the previous
publication~\cite{HIGG-2016-19}, the current expected limits are decreased 
by a factor
ranging from 20\% to 28\% for the ggF production mode and from 27\% to 43\% for the VBF production mode,
depending on the mass hypothesis.

\subsubsection{Spin-0 resonances with LWA}
\begin{figure}[htbp]
\centering
\subfigure[]{\includegraphics[width=0.45\textwidth]{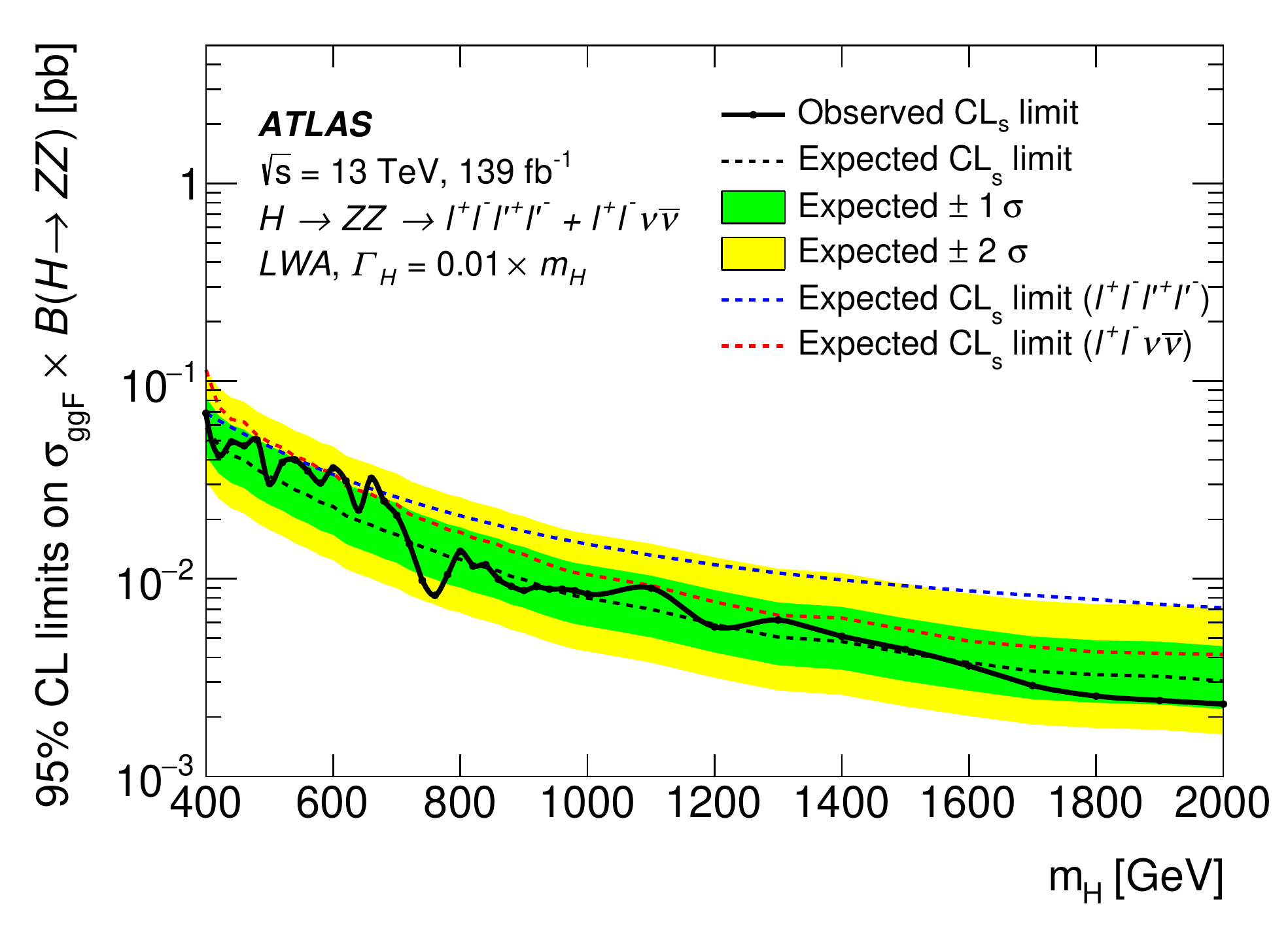}}
\subfigure[]{\includegraphics[width=0.45\textwidth]{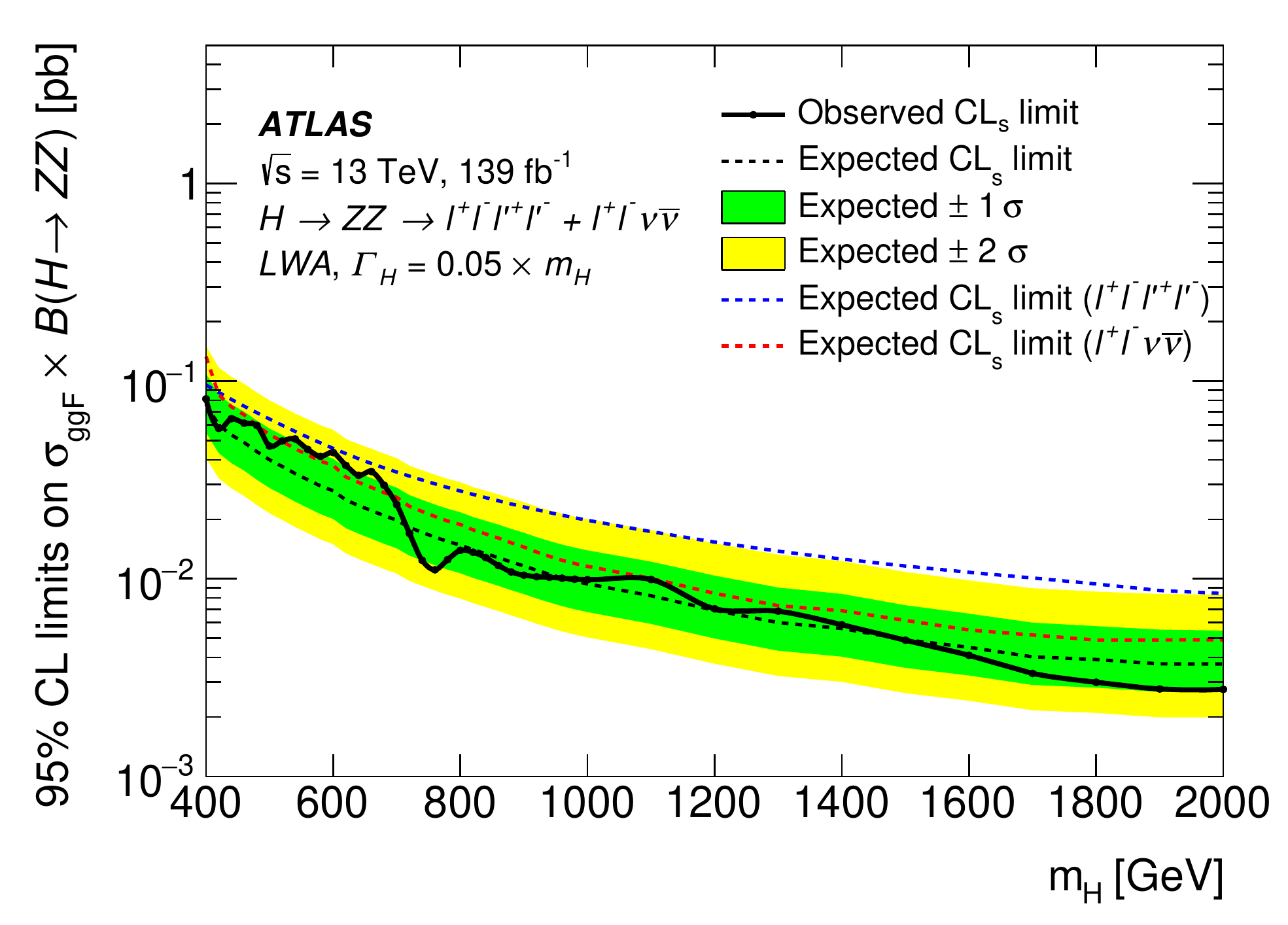}}
\subfigure[]{\includegraphics[width=0.45\textwidth]{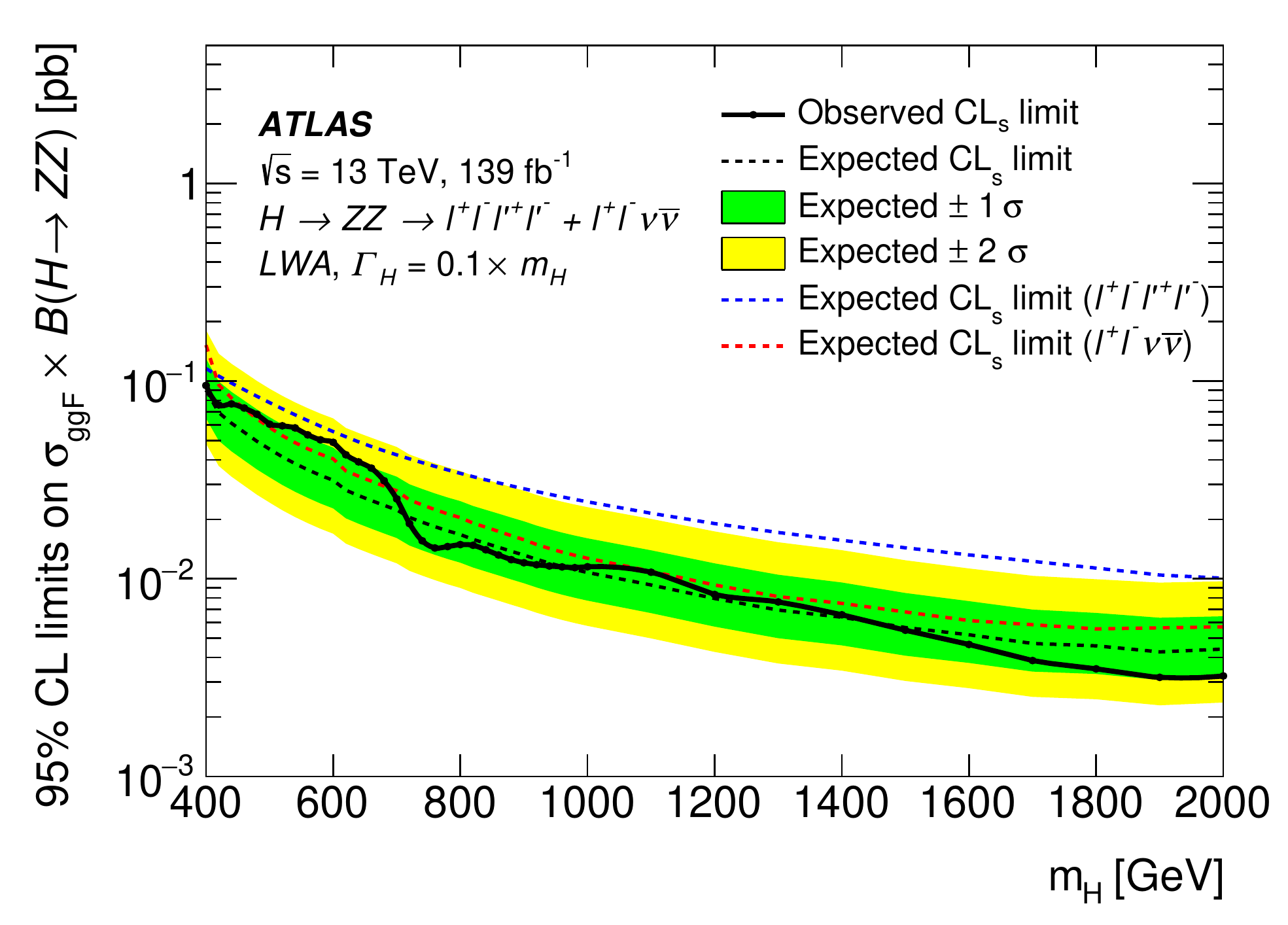}}
\subfigure[]{\includegraphics[width=0.45\textwidth]{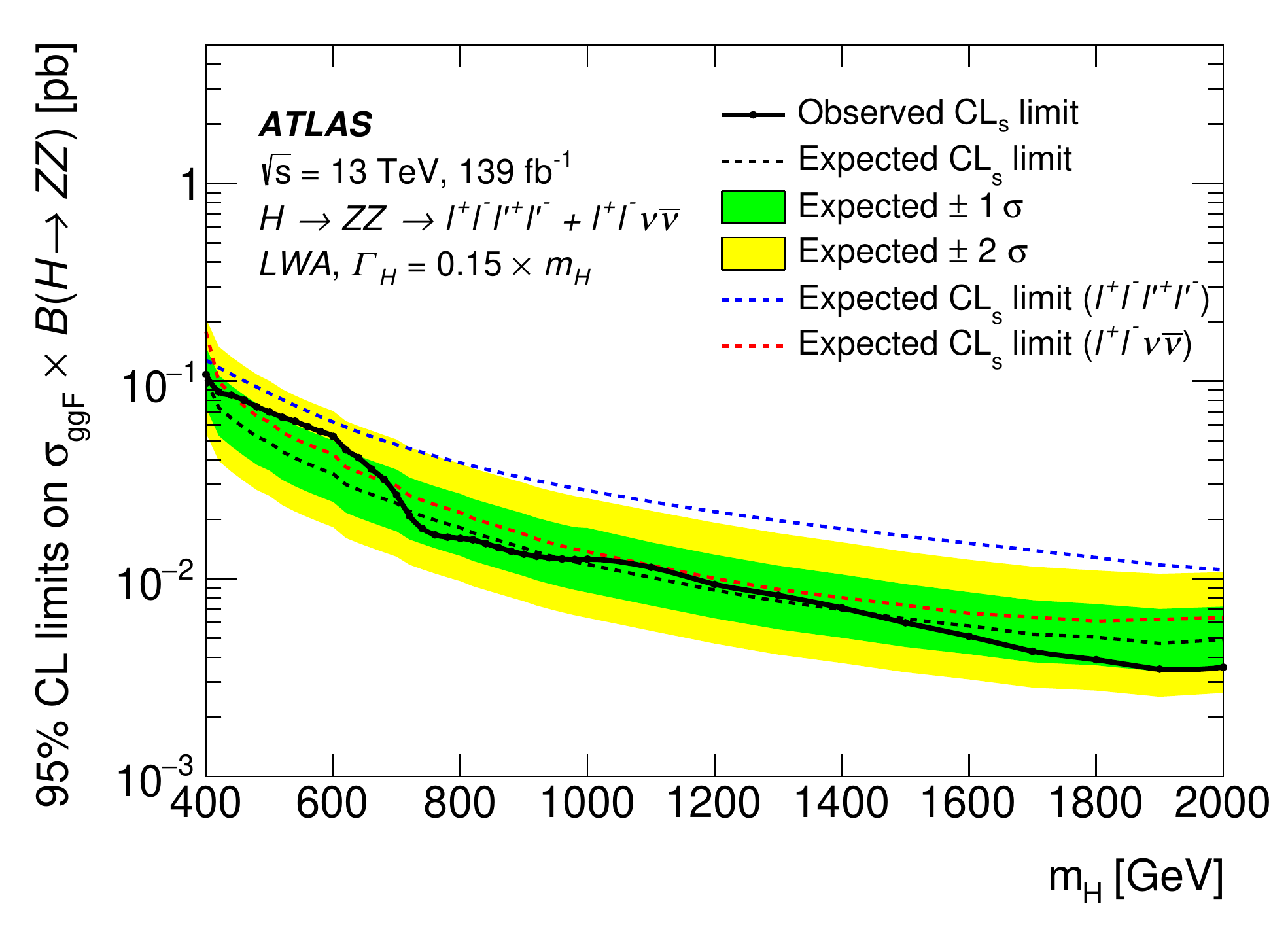}}
\caption{The upper limits at 95\% CL on the cross section for the
ggF production mode times branching ratio ($\sigma_{\textrm{ggF}}\times\mbr(\htozz)$)
as a function of \mres\ for an additional heavy scalar assuming
a width of (a) 1\%, (b) 5\%, (c) 10\% and (d) 15\%, of \mres.
The black line indicates the observed limit.
The green  and yellow  bands represent the $\pm$$1\sigma$ and $\pm$$2\sigma$
uncertainties in the expected limits.
The dashed coloured lines indicate the expected limits obtained from the individual searches.
}
\label{fig:LWAlimits}
\end{figure}

In the case of the LWA, upper limits on the cross section for the ggF process times branching ratio
($\xsggF \times \mbr (\htozz)$) are set for different widths of the heavy scalar.
Figures~\ref{fig:LWAlimits} shows the limits for a width of 1\%, 5\%, 10\% and 15\% of
\mres respectively. The limits are set for masses of \mres\ higher than $400~\gev{}$.
The choice of $400~\gev$ as the lower boundary is to avoid any major instability in the parametrization of
the mass spectra for the LWA signals and the interference effects, especially when
the signal pole mass gets smaller.
The interpretation has only been carried out for the ggF process,
as it is the leading channel to study the impact of non-trivial nature width on the search.

\subsubsection{Two-Higgs-doublet model}
A search in the context of a CP-conserving 2HDM is also
presented. This model has five physical Higgs bosons after electroweak symmetry breaking: two CP-even, one CP-odd, and two charged. The model considered here has seven free parameters: the
Higgs boson masses, the ratio of the vacuum expectation values of the two Higgs doublets ($\tan\beta$), the
mixing angle between the CP-even Higgs bosons ($\alpha$), and the potential parameter $m_{12}^2$ that mixes the two Higgs doublets.
The two Higgs doublets $\Phi_1$ and $\Phi_2$ can couple to
leptons and up- and down-type quarks in several ways.  In the Type-I model,
$\Phi_2$ couples to all quarks and leptons, whereas for Type-II,
$\Phi_1$ couples to down-type quarks and leptons and $\Phi_2$
couples to up-type quarks.  The `lepton-specific' model is similar to Type-I except for the fact
that the leptons couple to $\Phi_1$, instead of $\Phi_2$; the `flipped' model is similar to Type-II except that the leptons
couple to $\Phi_2$, instead of $\Phi_1$.
In all these models, the coupling of the heavier CP-even Higgs boson to vector bosons is proportional to
$\cos(\beta-\alpha)$. In the limit $\cos(\beta-\alpha) \rightarrow 0$, the
light CP-even Higgs boson is indistinguishable from a SM
Higgs~boson with the same mass. In the context of \htozz\ decays there is no direct coupling of the
Higgs~boson to leptons, so only the Type-I and II interpretations are
presented. In addition, our interpretations assume other Higgs bosons are heavy enough so that the heavy CP-even Higgs boson will not decay to them.
 
\begin{figure}[tbp]
\centering
\subfigure[]{\includegraphics[width=0.49\textwidth]{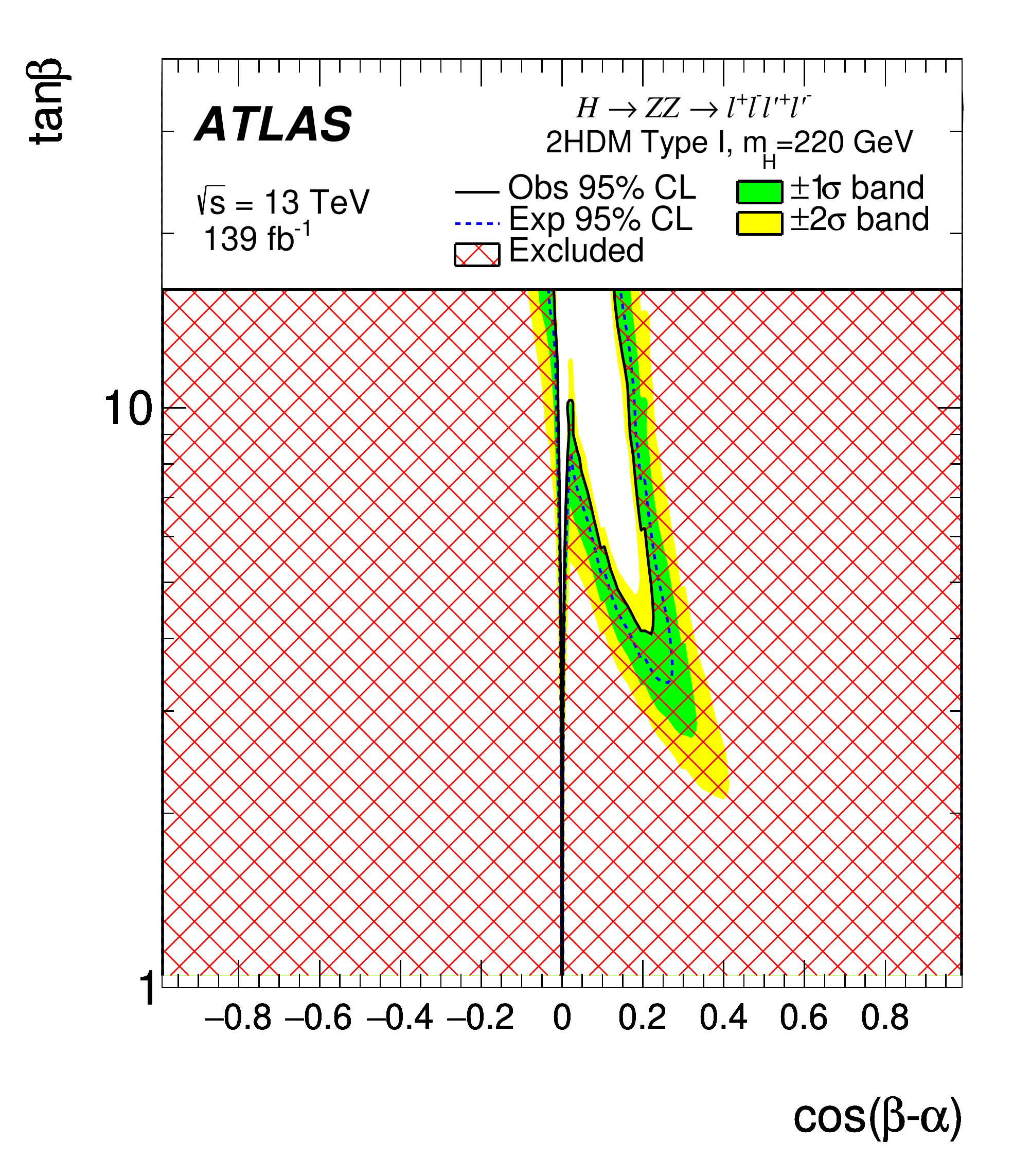}}
\subfigure[]{\includegraphics[width=0.49\textwidth]{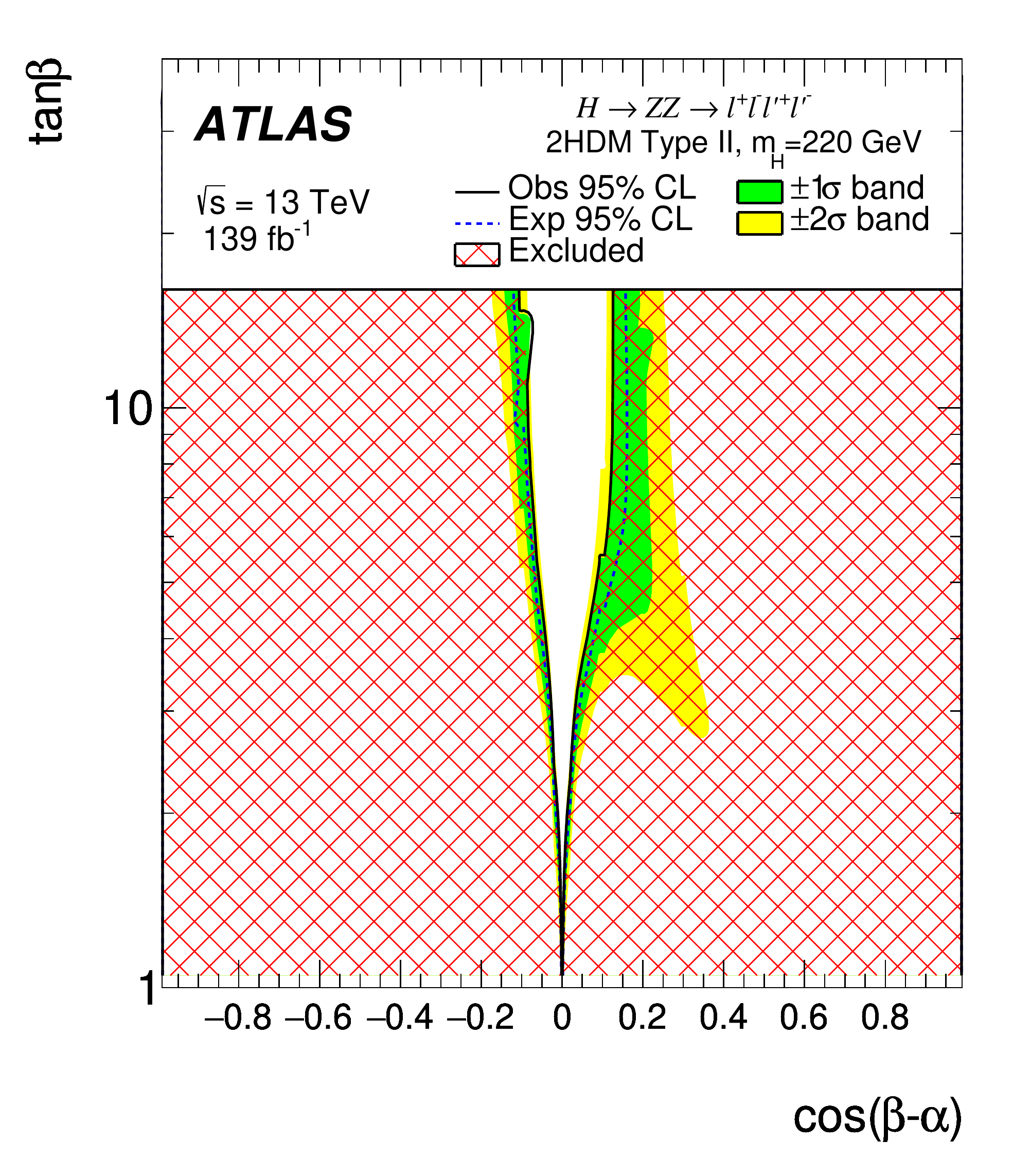}}
\caption{The exclusion contour in the 2HDM (a) Type-I
and (b) Type-II models for $\mH = 220~\gev{}$
shown as a function of the parameters $\cos(\beta-\alpha)$ and $\tan\beta$.
The green  and yellow  bands represent the $\pm$$1\sigma$ and $\pm$$2\sigma$
uncertainties in the expected limits. The hatched area shows the observed exclusion.
}
\label{fig:2HDM_tanb_cba_mH220}
\end{figure}

\begin{figure}[tbp]
\centering
\subfigure[]{\includegraphics[width=0.49\textwidth]{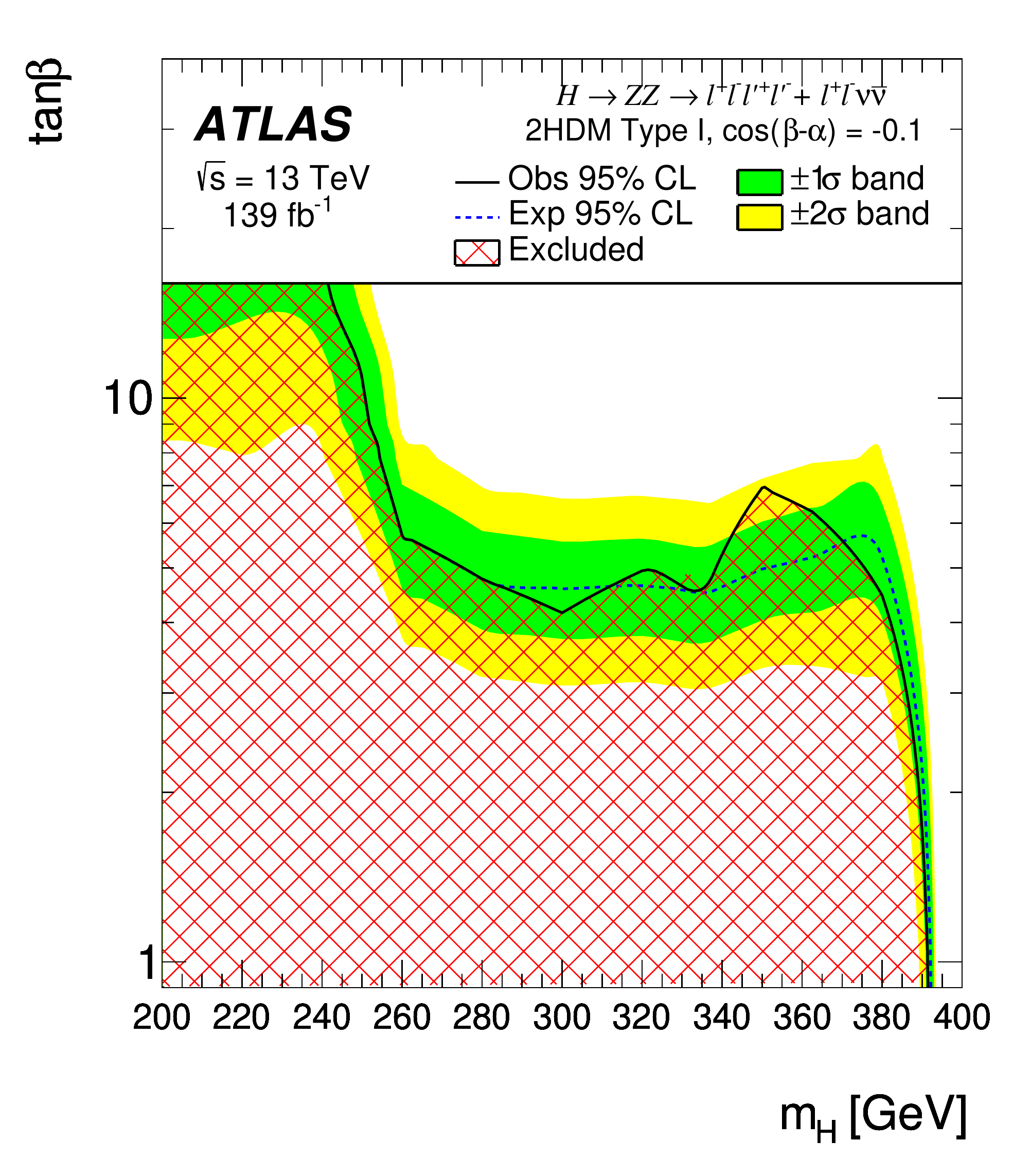}}
\subfigure[]{\includegraphics[width=0.49\textwidth]{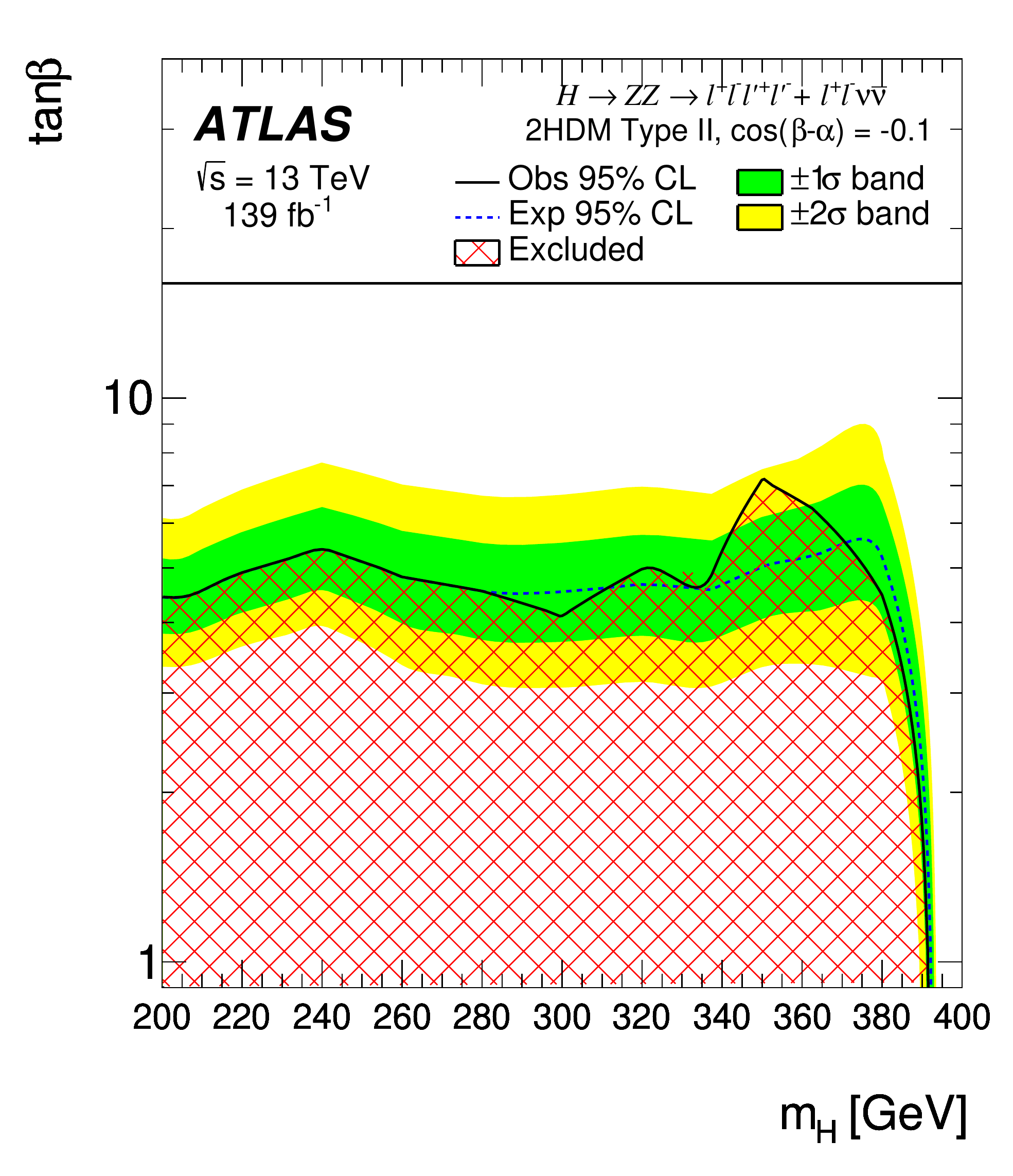}}
\caption{The exclusion contour in the 2HDM (a) Type-I
and (b) Type-II models for $\cos(\beta-\alpha)=-0.1$,
shown as a function of the heavy scalar mass \mres\ and the parameter $\tan\beta$.
The green  and yellow  bands represent the $\pm$$1\sigma$ and $\pm$$2\sigma$ uncertainties
in the expected limits. The hatched area shows the observed exclusion.
}
\label{fig:2HDM_tanb_mH}
\end{figure}

Figure~\ref{fig:2HDM_tanb_cba_mH220} shows exclusion limits in the $\tan\beta$ versus $\cos(\beta-\alpha)$
plane for Type-I and Type-II 2HDMs, for a heavy Higgs boson with mass $\mH =
220~\gev{}$. This \mH value is chosen so that the assumption of a narrow Higgs boson is valid
over most of the parameter space, and the experimental sensitivity is maximal. At this low mass,
only the \llll final state contributes to this result. The range of $\cos(\beta-\alpha)$ and
$\tan\beta$ explored is limited to the region where the assumption of a heavy narrow Higgs
boson with negligible interference is valid. When calculating the limits at a given choice of
$\cos(\beta-\alpha)$ and $\tan\beta$, the relative rates of ggF and VBF production in the fit are set
to the prediction of the 2HDM for that parameter choice. Figure~\ref{fig:2HDM_tanb_mH} shows
exclusion limits as a function of the heavy Higgs boson mass \mH and the parameter $\tan\beta$ for
$\cos(\beta-\alpha) = -0.1$, which is chosen so that the light Higgs boson properties are still compatible with the recent measurements of the SM Higgs boson properties~\cite{HIGG-2018-57}.
The white regions in the exclusion plots indicate regions of parameter
space which are not excluded by the present analysis. In these regions the cross section predicted
by the 2HDM is below the observed cross-section limit.
In comparison with the previous publication, the excluded regions are significantly expanded.
For example, in the \tanb\ versus \mH plane for the Type-II 2HDM the excluded region in \tanb\
is more than 60\% larger for $200 < \mH < 400~\gev{}$.

\subsection{Spin-2 resonances}
 
\begin{figure}[tbp]
\centering
\includegraphics[width=0.7\linewidth]{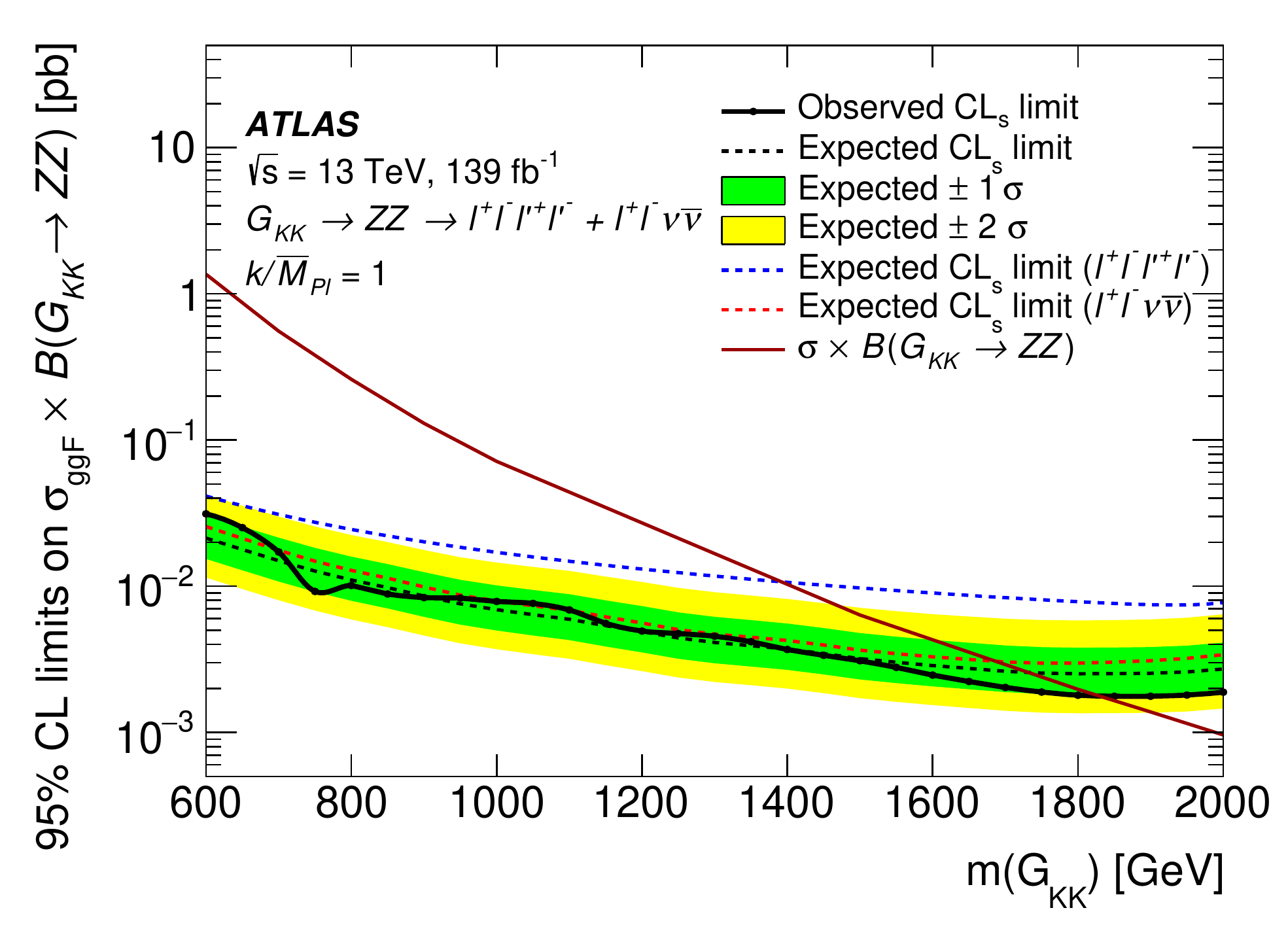}
\caption{The upper limits at 95\% CL on cross section times branching ratio
$\sigma \times\mbr(\Gratozz)$ for a KK graviton produced with $k/{\overline{M}_\mathrm{Pl}} = 1$.
The black line indicates the observed limit.
The green and yellow bands give the $\pm$$1\sigma$ and $\pm$$2\sigma$ uncertainties in the expected limits.
The predicted production cross section times branching ratio
as a function of the \grav mass \mgrav\ is shown by the red solid line.}
\label{fig:graviton}
\end{figure}

The results are also interpreted as a search for a Kaluza--Klein graviton excitation, \grav,
in the context of the bulk RS model with $k/{\overline{M}_\mathrm{Pl}} = 1$.
The limits on $\sigma \times\mbr(\Gratozz)$ at 95\% CL
as a function of the KK graviton mass, \mgrav, are shown in Figure~\ref{fig:graviton}
together with the predicted \grav\ cross section.
A spin-2 graviton is excluded up to a mass of $1830~\gev{}$.
 
\FloatBarrier
 
\section{Summary}
A search is conducted for heavy resonances decaying into a pair of \Z bosons which subsequently decay
into \llll or \llvv final states.
The search uses proton--proton collision data collected with the ATLAS detector from 2015 to 2018 at the Large Hadron
Collider at a centre-of-mass energy of 13~\tev\
corresponding to the full \RunTwo integrated luminosity of \lumi.
No significant excess is observed with respect to the predicted SM background; therefore, the results are interpreted as
upper limits on the production cross section of spin-0 resonances or a spin-2 resonance.
The mass range of the hypothetical resonances considered
is between $200~\gev{}$ and $2000~\gev{}$ depending on the final state and the model considered.
The spin-0 resonance is assumed to be a heavy scalar,
whose dominant production modes are gluon--gluon fusion and vector-boson fusion, and it is studied
in the narrow-width approximation and with the large-width assumption.
In the case of the narrow-width approximation, upper limits on the production rate of a heavy scalar decaying into two $Z$ bosons (the production cross-section times the corresponding decay branching fraction) are set separately for ggF and VBF production modes.
Combining the two final states, 95\% CL upper limits
range from \mbox{215~fb at $\mres = 240~\gev{}$ to 2.0~fb at $\mres = 1900~\gev{}$}
for the gluon--gluon fusion production mode
and from \mbox{87~fb at $\mres = 255~\gev{}$ to 1.5~fb at $\mres = 1800~\gev{}$} for the vector-boson fusion production mode.
The results are also interpreted in the context of Type-I and Type-II two-Higgs-doublet
models, with exclusion contours given in the \tanb\ versus \cosab\ (for $\mH = 220~\gev{}$) and \tanb\ versus \mH\ planes.
This \mH value is chosen so that the assumption of a narrow Higgs boson is
valid over most of the parameter space and the experimental sensitivity is maximal.  The limits on
the production rate of a large-width scalar are obtained for widths of 1\%, 5\%, 10\% and 15\% of the
mass of the resonance, with the interference between the heavy scalar and the SM Higgs boson as well
as between the heavy scalar and the \ggZZ continuum taken into account.
In the framework of the Randall--Sundrum model
with one warped extra dimension a graviton excitation spin-2 resonance
with $\mgrav < 1830~\gev{}$ is excluded at 95\% CL.

\section*{Acknowledgements}


We thank CERN for the very successful operation of the LHC, as well as the
support staff from our institutions without whom ATLAS could not be
operated efficiently.
 
We acknowledge the support of
ANPCyT, Argentina;
YerPhI, Armenia;
ARC, Australia;
BMWFW and FWF, Austria;
ANAS, Azerbaijan;
SSTC, Belarus;
CNPq and FAPESP, Brazil;
NSERC, NRC and CFI, Canada;
CERN;
ANID, Chile;
CAS, MOST and NSFC, China;
Minciencias, Colombia;
MEYS CR, Czech Republic;
DNRF and DNSRC, Denmark;
IN2P3-CNRS and CEA-DRF/IRFU, France;
SRNSFG, Georgia;
BMBF, HGF and MPG, Germany;
GSRI, Greece;
RGC and Hong Kong SAR, China;
ISF and Benoziyo Center, Israel;
INFN, Italy;
MEXT and JSPS, Japan;
CNRST, Morocco;
NWO, Netherlands;
RCN, Norway;
MEiN, Poland;
FCT, Portugal;
MNE/IFA, Romania;
JINR;
MES of Russia and NRC KI, Russian Federation;
MESTD, Serbia;
MSSR, Slovakia;
ARRS and MIZ\v{S}, Slovenia;
DSI/NRF, South Africa;
MICINN, Spain;
SRC and Wallenberg Foundation, Sweden;
SERI, SNSF and Cantons of Bern and Geneva, Switzerland;
MOST, Taiwan;
TAEK, Turkey;
STFC, United Kingdom;
DOE and NSF, United States of America.
In addition, individual groups and members have received support from
BCKDF, CANARIE, Compute Canada and CRC, Canada;
COST, ERC, ERDF, Horizon 2020 and Marie Sk{\l}odowska-Curie Actions, European Union;
Investissements d'Avenir Labex, Investissements d'Avenir Idex and ANR, France;
DFG and AvH Foundation, Germany;
Herakleitos, Thales and Aristeia programmes co-financed by EU-ESF and the Greek NSRF, Greece;
BSF-NSF and GIF, Israel;
Norwegian Financial Mechanism 2014-2021, Norway;
NCN and NAWA, Poland;
La Caixa Banking Foundation, CERCA Programme Generalitat de Catalunya and PROMETEO and GenT Programmes Generalitat Valenciana, Spain;
G\"{o}ran Gustafssons Stiftelse, Sweden;
The Royal Society and Leverhulme Trust, United Kingdom.
 
The crucial computing support from all WLCG partners is acknowledged gratefully, in particular from CERN, the ATLAS Tier-1 facilities at TRIUMF (Canada), NDGF (Denmark, Norway, Sweden), CC-IN2P3 (France), KIT/GridKA (Germany), INFN-CNAF (Italy), NL-T1 (Netherlands), PIC (Spain), ASGC (Taiwan), RAL (UK) and BNL (USA), the Tier-2 facilities worldwide and large non-WLCG resource providers. Major contributors of computing resources are listed in Ref.~\cite{ATL-SOFT-PUB-2021-003}.
 

\clearpage
\part*{Appendix}
\begin{figure}[tbp]
 
\subfigure[]{\includegraphics[width=0.49\textwidth]{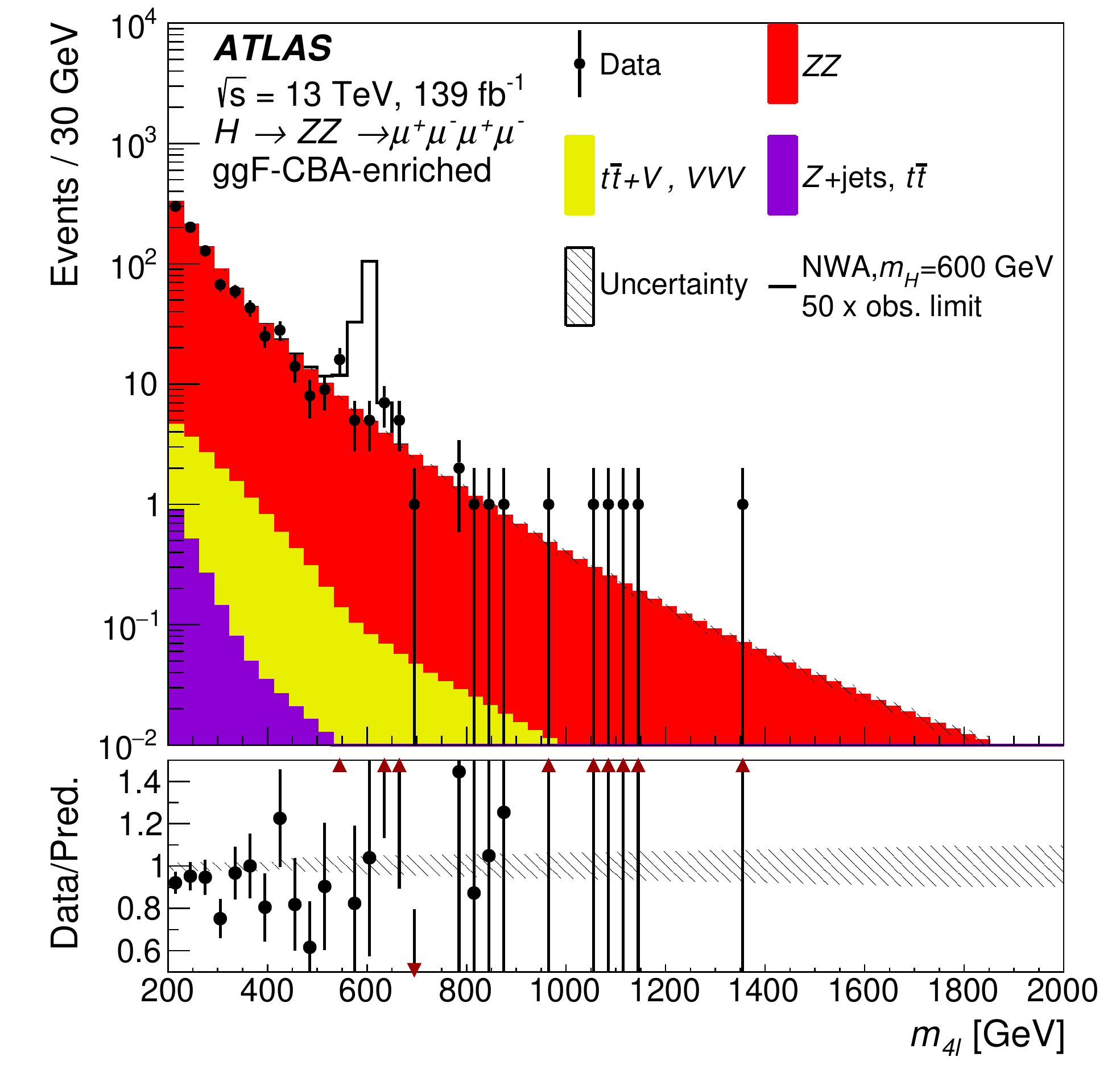}}
\subfigure[]{\includegraphics[width=0.49\textwidth]{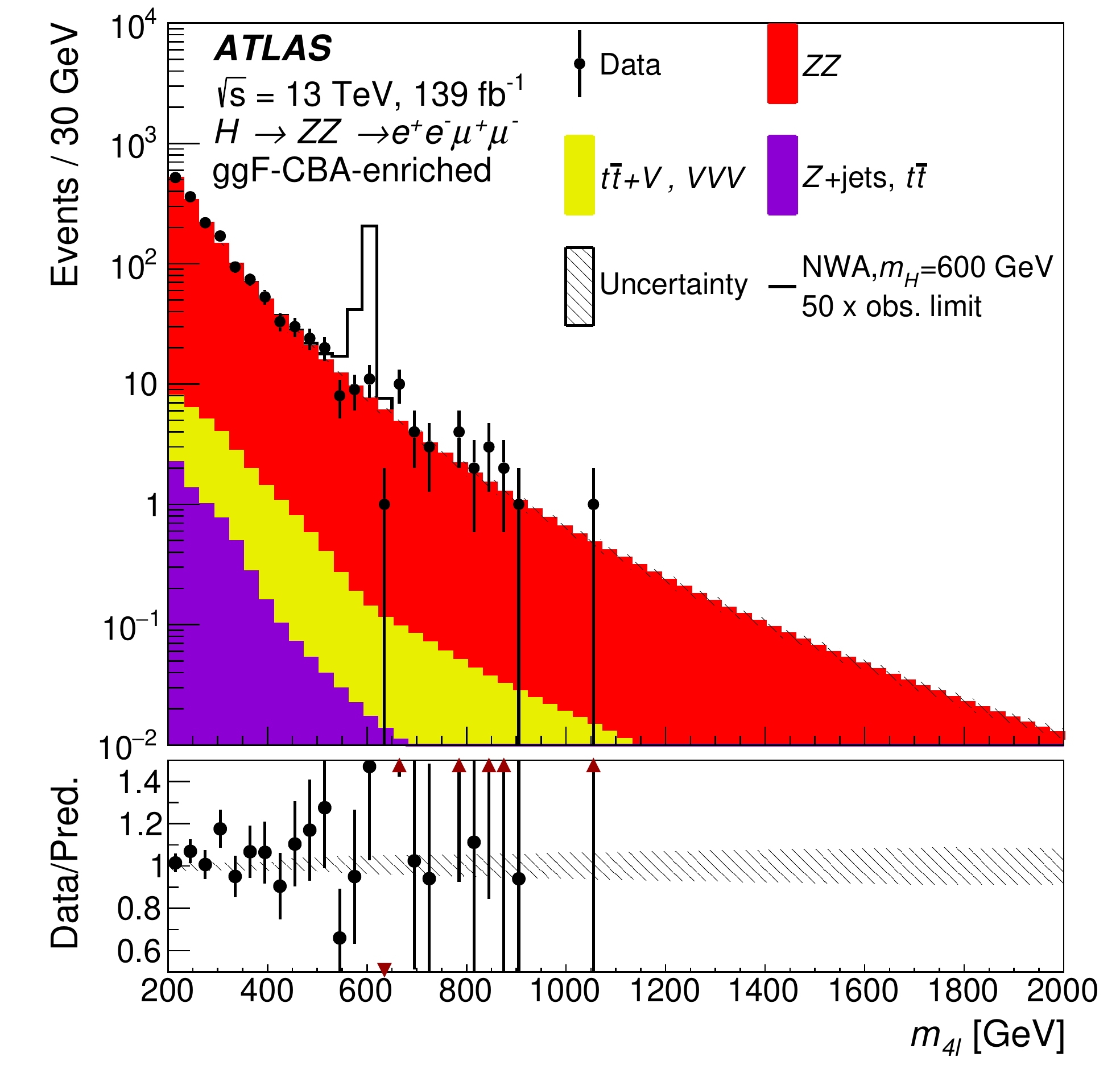}}
\subfigure[]{\includegraphics[width=0.49\textwidth]{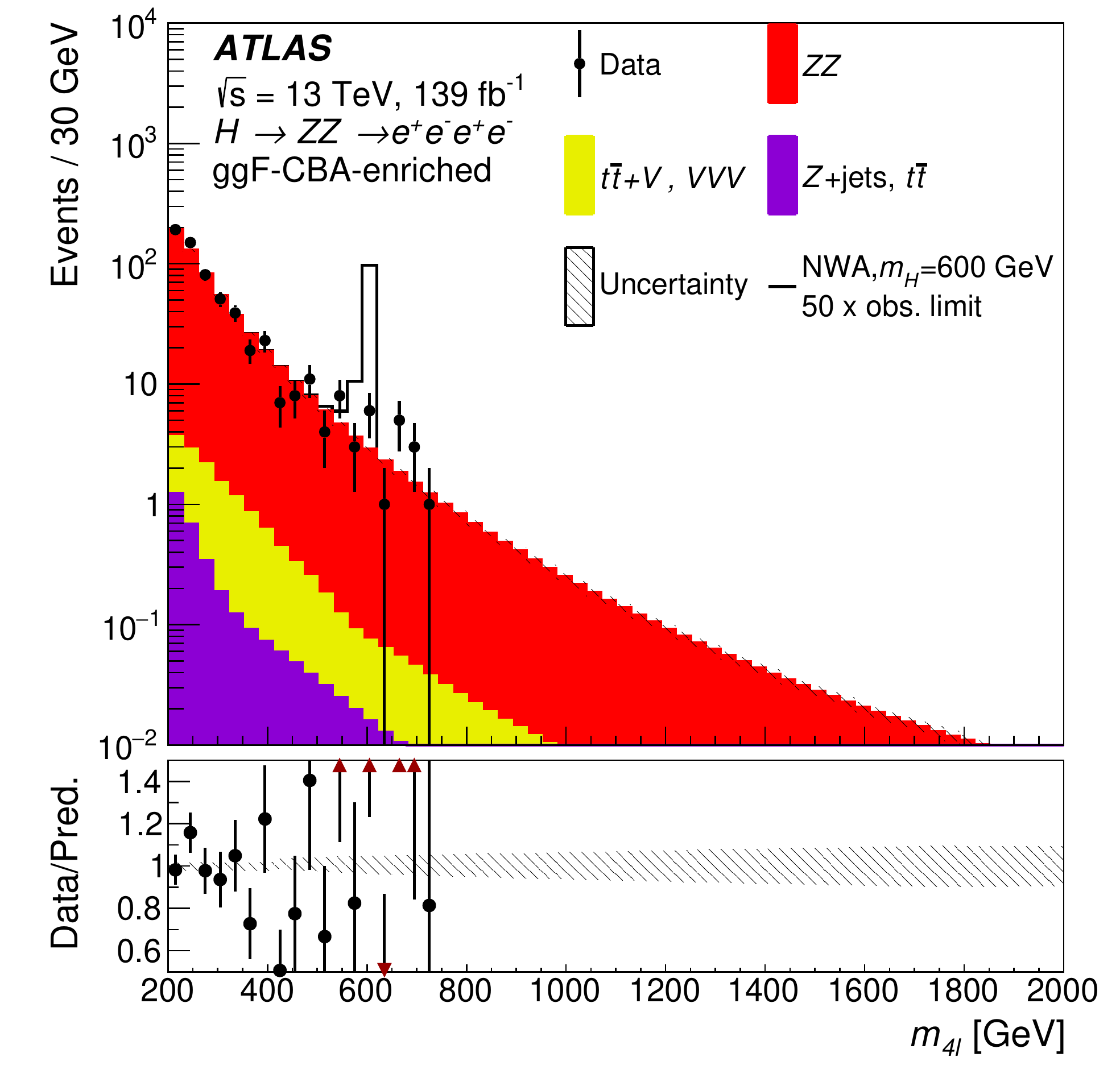}}
\subfigure[]{\includegraphics[width=0.49\textwidth]{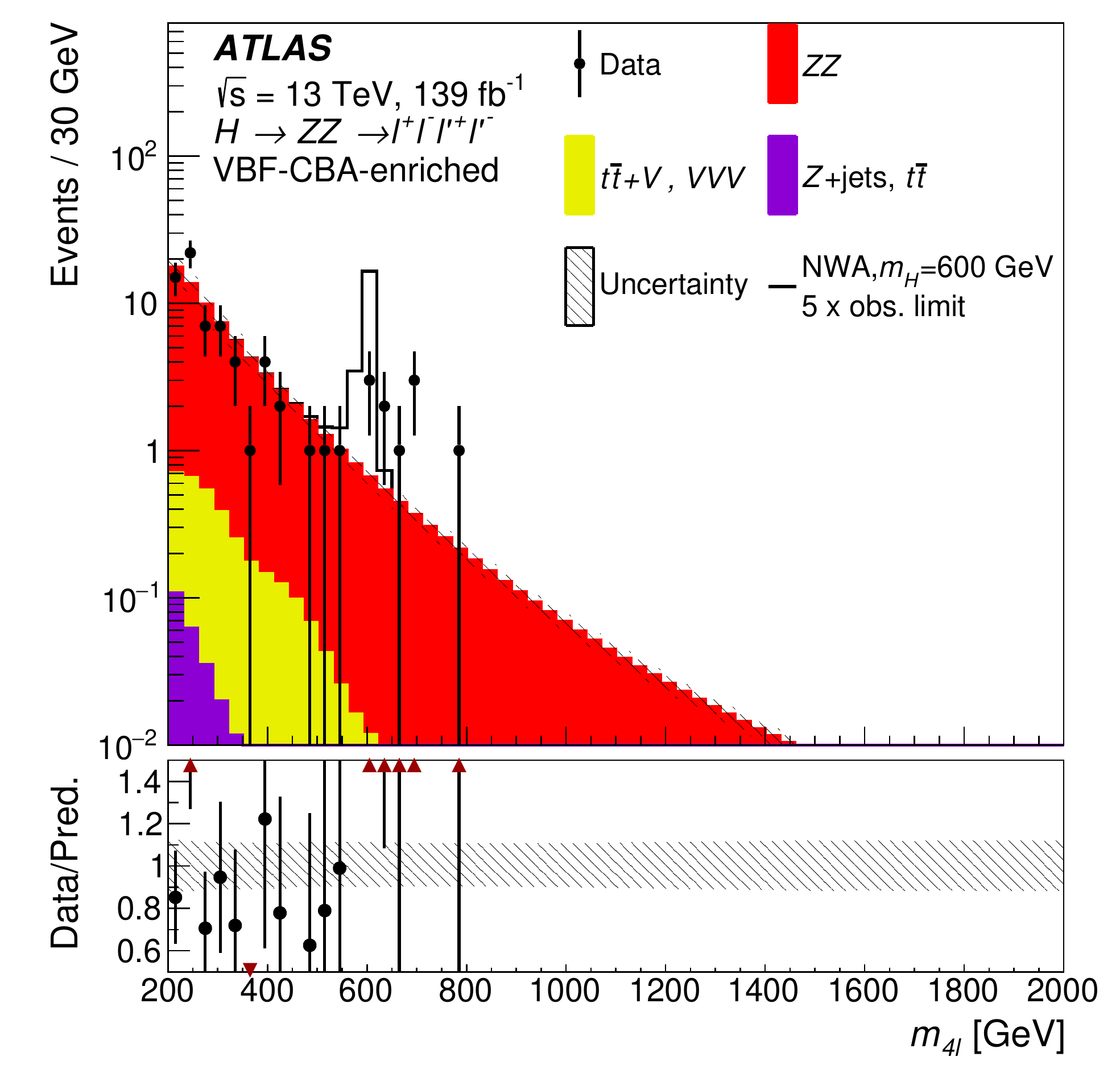}}
 
\caption{Distribution of the four-lepton invariant mass \mfl in the $\llll$ search
for  (a),(b),(c) the \ggfCBAenr categories
and (d) the \vbfCBAenr category.
The backgrounds are determined from a combined likelihood fit to the data under the background-only
hypothesis. The simulated $\mres = 600~\gev{}$ signal is normalised to
a cross section corresponding to 50 (5) times the observed limit given in Section~\ref{sec:spin0nwa} for the ggF (VBF) production mode. The error bars on the data points indicate the statistical uncertainty,
while the systematic uncertainty in the prediction is shown by the \systband\ band.
The lower panels show the ratio of data to prediction.}
\label{fig:m4l_sig-cut}
\end{figure}

\begin{figure}[htbp]
\subfigure[]{\includegraphics[width=0.49\textwidth]{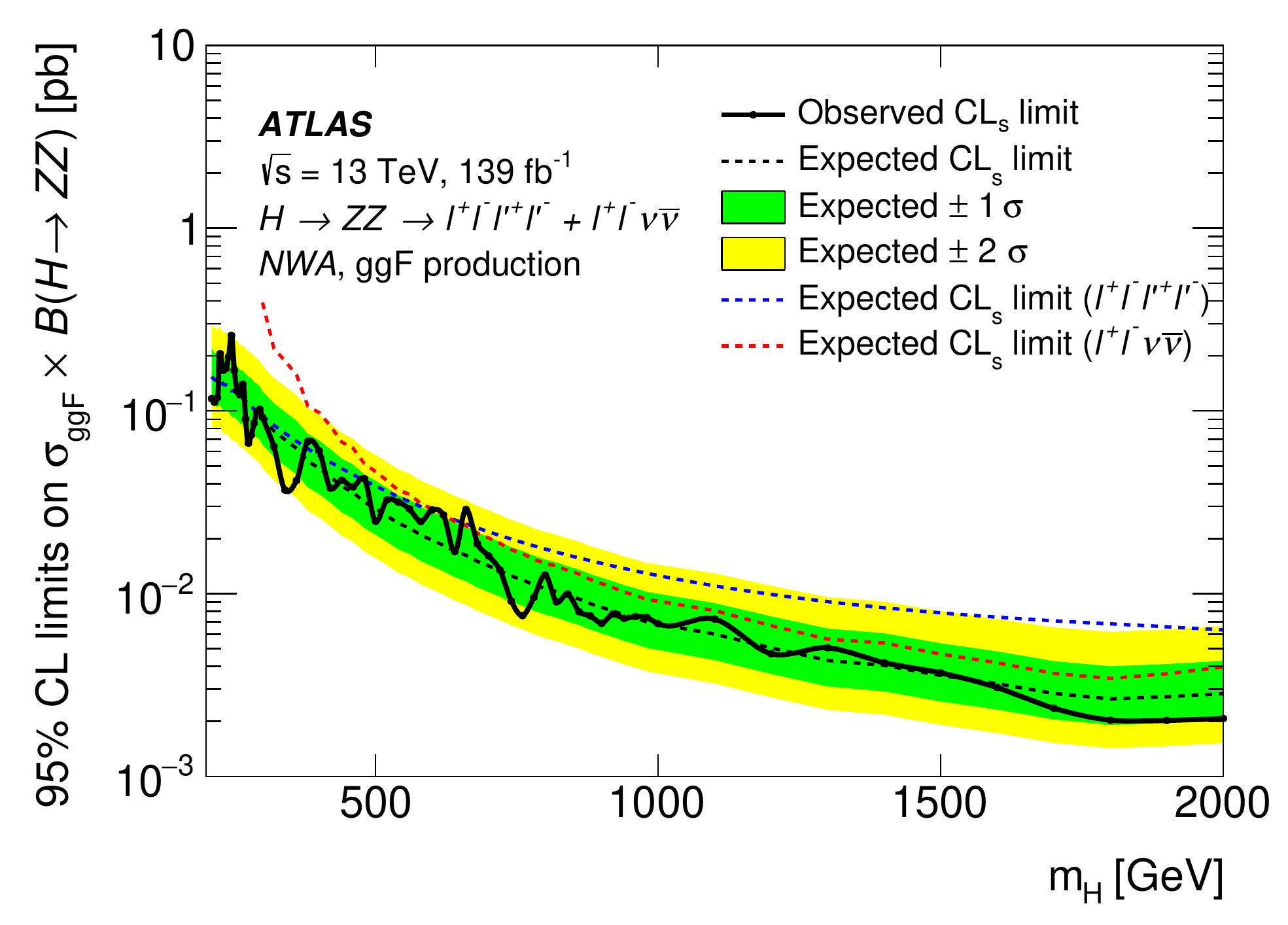}}
\subfigure[]{\includegraphics[width=0.49\textwidth]{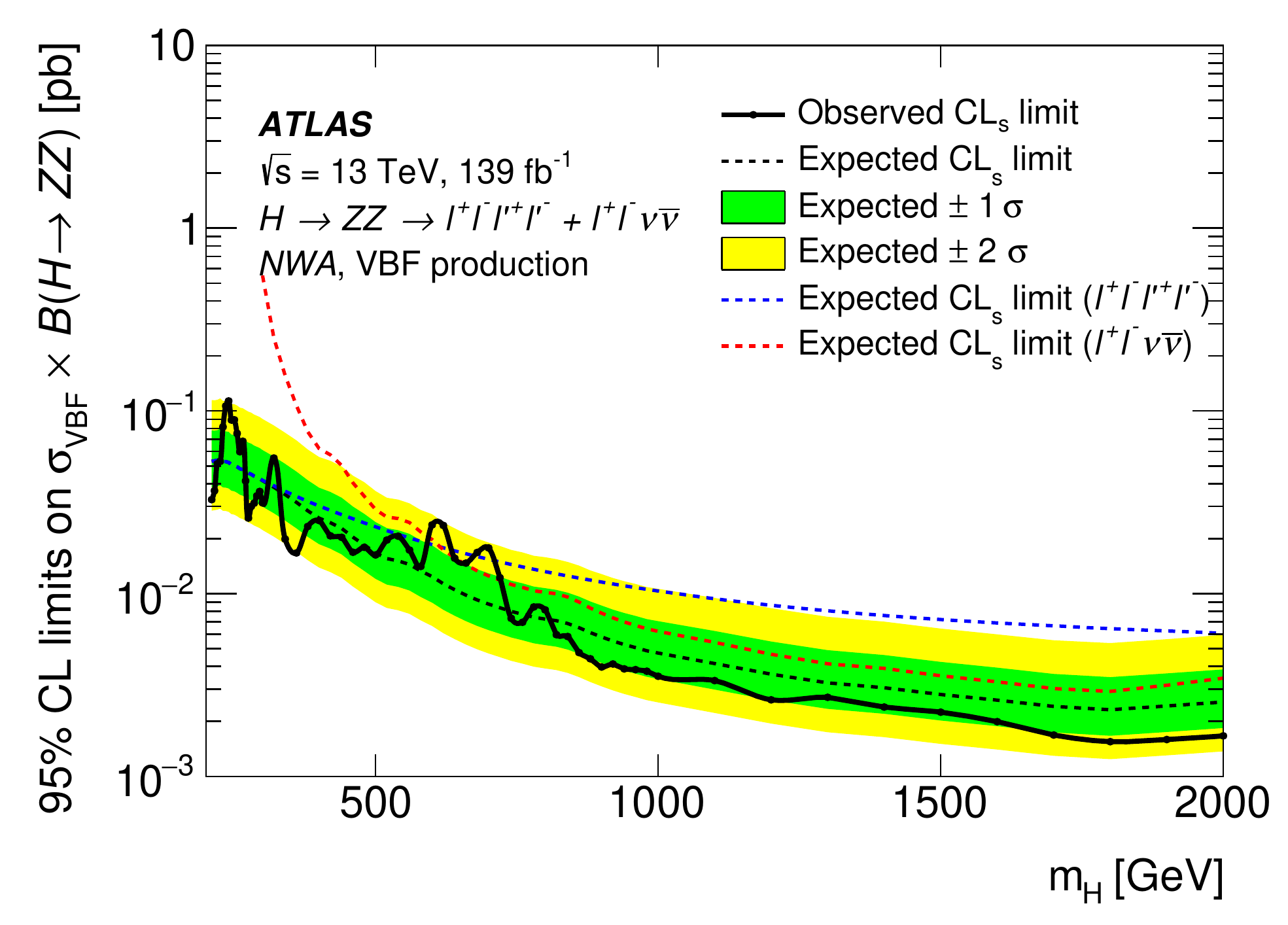}}
\caption{The upper limits at 95\% CL on the cross section times branching ratio
as a function of the heavy resonance mass $\mres$ for (a)
the ggF production mode ($\xsggF\times\mbr(\htozz)$) and (b)
for the VBF production mode  ($\xsVBF\times\mbr(\htozz)$)  in the case of the NWA.
For the \llll state the cut-based categorisation is used.
The green and yellow  bands represent the $\pm$$1\sigma$ and $\pm$$2\sigma$ uncertainties
in the expected limits. The dashed coloured lines indicate the expected limits obtained
from the individual searches.
}
\label{fig:nwa-limits-cutbased}
\end{figure}

\addcontentsline{toc}{part}{Appendix}
The results based on the cut-based categorisation as described in Section~\ref{sec:vbfsel4l} are presented here.
The number of observed candidate events with mass above $200~\gev{}$ together with the expected background yields for each of the four categories of the \llll analysis as described in Section~\ref{sec:vbfsel4l} is presented in Table~\ref{tab:H4lyields-cut}.
The obtained \ZZ normalisation factors are summarised in Table~\ref{tab:fittedFactors-cutbased}, and the \mfl spectrum in each category is shown in Figure~\ref{fig:m4l_sig-cut}.
The upper limits at 95\% CL on the cross section times branching ratio as a function of the heavy resonance mass in the case of the NWA is presented in Figure~\ref{fig:nwa-limits-cutbased}.

\begin{table}[htbp]
\caption{
\llll search: expected and observed numbers of events for $\mfl > 200~\gev{}$,
together with their uncertainties, for the \vbfCBAenr and \ggfCBAenr categories using
the cut-based categorisation.
The expected numbers of events, as well as their uncertainties, are obtained
from a combined likelihood fit to the data under the background-only hypothesis.
The uncertainties of the \ZZ normalisation factors,
presented in Table~\ref{tab:fittedFactors-cutbased},
are also taken into account.
}
\label{tab:H4lyields-cut}
\centering
 
\tiny
\resizebox{\textwidth}{!}{
\begin{tabular}{l
S[table-format=2.1]  @{ $\pm$ } S[table-format=1.1]
S[table-format=3.1]  @{ $\pm$ } S[table-format=2.1]
S[table-format=4.1]  @{ $\pm$ } S[table-format=2.1]
S[table-format=3.1]  @{ $\pm$ } S[table-format=2.1]
}
\toprule
\multirow{2}{*}{Process} & \multicolumn{2}{c}{\vbfCBAenr} & \multicolumn{6}{c}{\ggfCBAenr} \\
& \multicolumn{2}{c}{} & \multicolumn{2}{c}{$4\mu$ channel} & \multicolumn{2}{c}{$2e2\mu$ channel} & \multicolumn{2}{c}{$4e$ channel} \\
\midrule
 
\qqZZ	&	48	&	8	&	860	&	18	&	1360	&	28	&	515	&	11	\\
\ggZZ	&	13	&	4	&	114	&	9	&	189	&	14	&	73	&	6	\\
\ZZ (EW)	&	10.9	&	0.9	&	6.9	&	0.3	&	11.1	&	0.4	&	4.4	&	0.2	\\
\Zjets/\ttbar/\WZ	&	0.3	&	0.1	&	2.1	&	0.4	&	6.7	&	1.6	&	3.1	&	0.4	\\
Other backgrounds	&	3	&	0.2	&	16.3	&	0.4	&	26.8	&	0.6	&	11.8	&	0.3	\\
\midrule
Total background	&	75	&	9	&	999	&	20	&	1594	&	31	&	607	&	13	\\

\midrule
Observed  &\multicolumn{2}{l}{75} & \multicolumn{2}{l}{932} & \multicolumn{2}{l}{1656} & \multicolumn{2}{l}{612} \\
\bottomrule
\end{tabular}
}
\end{table}

\begin{table}[htb]
\caption{
The \ZZ normalisation factors in each category of the two final states,
which scale the number of \ZZ events estimated from the MC simulations,
obtained from a simultaneous likelihood fit of the two final states
under the background-only hypothesis.
For the \llll final state, the cut-based categorisation is used.
}
\label{tab:fittedFactors-cutbased}
\centering
\begingroup
\setlength{\tabcolsep}{10pt}
\renewcommand{\arraystretch}{1.5}
\begin{tabular}{c|
c|
S[table-format=1.2]  @{ $\pm$ } S[table-format=1.2]  }
\hline
Analysis & Normalisation factor & \multicolumn{2}{c}{Fitted value} \\ \hline
\multirow{2}{*}{\llll} & $\mu_{ZZ}^{\text{VBF-CBA}}$    &  1.1  & 0.2  \\
& \muZZggF  &  1.10 & 0.02 \\
\midrule
\llvv                  & \muZZlv     &  1.04& 0.06 \\ \hline
\end{tabular}
\endgroup
\end{table}
\clearpage
 
\printbibliography

\clearpage
 
\begin{flushleft}
\hypersetup{urlcolor=black}
{\Large The ATLAS Collaboration}

\bigskip

\AtlasOrcid[0000-0002-6665-4934]{G.~Aad}$^\textrm{\scriptsize 102}$,    
\AtlasOrcid[0000-0002-5888-2734]{B.~Abbott}$^\textrm{\scriptsize 128}$,    
\AtlasOrcid[0000-0002-7248-3203]{D.C.~Abbott}$^\textrm{\scriptsize 103}$,    
\AtlasOrcid[0000-0002-2788-3822]{A.~Abed~Abud}$^\textrm{\scriptsize 36}$,    
\AtlasOrcid[0000-0002-1002-1652]{K.~Abeling}$^\textrm{\scriptsize 53}$,    
\AtlasOrcid[0000-0002-2987-4006]{D.K.~Abhayasinghe}$^\textrm{\scriptsize 94}$,    
\AtlasOrcid[0000-0002-8496-9294]{S.H.~Abidi}$^\textrm{\scriptsize 167}$,    
\AtlasOrcid[0000-0002-8279-9324]{O.S.~AbouZeid}$^\textrm{\scriptsize 40}$,    
\AtlasOrcid{N.L.~Abraham}$^\textrm{\scriptsize 156}$,    
\AtlasOrcid[0000-0001-5329-6640]{H.~Abramowicz}$^\textrm{\scriptsize 161}$
% ...
\bigskip
\\

% ...
$^{36}$CERN, Geneva; Switzerland.\\
% ...
$^{40}$Niels Bohr Institute, University of Copenhagen, Copenhagen; Denmark.\\
% ...
$^{53}$II. Physikalisches Institut, Georg-August-Universit\"{a}t G\"ottingen, G\"ottingen; Germany.\\
% ...
$^{94}$Department of Physics, Royal Holloway University of London, Egham; United Kingdom.\\
% ...
$^{102}$CPPM, Aix-Marseille Universit\'e, CNRS/IN2P3, Marseille; France.\\
$^{103}$Department of Physics, University of Massachusetts, Amherst MA; United States of America.\\
% ...
$^{128}$Homer L. Dodge Department of Physics and Astronomy, University of Oklahoma, Norman OK; United States of America.\\
% ...
$^{156}$Department of Physics and Astronomy, University of Sussex, Brighton; United Kingdom.\\
% ...
$^{161}$Raymond and Beverly Sackler School of Physics and Astronomy, Tel Aviv University, Tel Aviv; Israel.\\
% ...
$^{167}$Department of Physics, University of Toronto, Toronto ON; Canada.\\
% ...

\end{flushleft}

 
\end{document}